\documentclass{article}
\usepackage{graphicx}
\usepackage{amssymb}
\usepackage{amsmath}
\usepackage{comment}

\begin{document}

\title{{\bf Internal quark symmetries and colour $SU(3)$ entangled with $Z_3$-graded Lorentz algebra}}

\author{Richard Kerner$^a$ and Jerzy Lukierski$^{b}$ 
\\ \\ 
{\small  a : Laboratoire de Physique Th\'eorique de la Mati\`ere}  \\
{\small Condens\'ee - CNRS URA 7600 - Sorbonne-Universit\'e, B.C. 121,} 
 \\
{\small 4 Place Jussieu 75005 Paris, France}
\\
{\small b: Institute of Theoretical Physics, Wroc{\l}aw University, }
 \\ 
{\small Plac Maxa Borna 9,  Wroc{\l}aw, Poland }
 \\
{\small e-mail  a: richard.kerner@sorbonne-universite.fr}
\\
{\small e-mail  b: jerzy.lukierski@ift.uni.wroc.pl} }

\maketitle

\begin{abstract}

In the current version of QCD the quarks are described by ordinary Dirac fields, organized in the following
internal symmetry multiplets: the $SU(3)$ colour, the $SU(2)$ flavour, and broken $SU(3)$ providing the family triplets.
\noindent
In this paper we argue that internal and external (i.e. space-time) symmetries are entangled at least in the colour sector in order
to introduce the spinorial quark fields in a way providing all the internal quark's degrees of freedom which do appear in the Standard Model. 
Because the $SU(3)$ colour algebra is endowed with natural $Z_3$-graded discrete automorphisms, in order to introduce entanglement the
$Z_3$-graded version of Lorentz algebra with its vectorial and spinorial realizations are considered.  
%We show how the internal symmetry $SU(3)_{colour}$ appears naturally as the group of invariance of the $Z_3$-graded Lorentz group.
The colour multiplets of quarks are described by $12$-component colour Dirac equations, with a $Z_3$-graded triplet of masses
(one real and a Lee-Wick complex conjugate pair). We show that all quarks in the Standard Model can be described by the $72$-component 
master quark sextet of $12$-component coloured Dirac fields, which is required in order to implement the faithful spinorial representation
of the $Z_3$-graded Lorentz transformations.

\end{abstract}

\section{Introduction}
\vskip 0.2cm
\indent
In the current version of Quantum Chromodynamics the massive quarks are treated as Dirac fermions endowed
with additional internal degrees of freedom. In the minimal version, Standard Model displays the exact $SU(3)$ colour 
and the $SU(2)$ flavour symmetries, as well as strongly broken $SU(3)$ describing three quark families (\cite{Gaillard}, \cite{Cottingham},
 \cite{Bustamante}).

If we introduce ``master quark Dirac field'' supposed to incorporate all internal quark symmetries, we should
deal with $4 \times 3 \times 2 \times 3 = 72$-component fermionic master field (the first factor $4$ corresponds to the
four degrees of freedom of classical Dirac spinor, the next factor $3$ stays for three colours, next factor $2$ gives
 flavours, and the last factor $3$ corresponds to the three families). Our aim here is to look for a framework introducing 
algebraic and group-theoretical structure which permits to incorporate all the internal 
quark symmetries enumerated above in some irreducible representations of $Z_3$-graded generalization of Lorentz algebra.

Because in the quark sector of Standard Model e.g. $u$ and $d$ quarks behave as fermions, the two-quark states $uu$ or $dd$ should be excluded, 
unless there are extra parameters distinguishing the states in a pair. This was the origin of {\it colour} degrees of freedom,
and of an exact $SU(3)$ colour symmetry treating quarks as colour triplets which incorporate
{\it three} distinct eigenstates, labeled as {\it red, green,} and {\it blue}. With such
enlargement of the Hilbert space describing single quark states we arrive in Sect. $3$ at new $12$-component fermionic colour Dirac field, 
introduced in \cite{RKOS2014}, \cite{Kerner2017A}, which is covariant under $Z_3$ symmetry group and contains colour Dirac $\Gamma^{\mu}$ 
matrices whose structure is described symbolically by the following tensor product:
\begin{equation}
M_3 (C) \otimes H_2 \otimes H_2
\label{sixferm}
\end{equation}
$M_3 (C)$ represents colour $3 \times 3$ matrices, $H_2$ are the $2 \times 2$ Hermitiean ones, and in our approach
the generalized $12 \times 12$-dimensional generalized Dirac $\Gamma^{\mu}$-matrices employ in $M_3$ sector the generators of the
particular ternary Clifford algebra discussed in Sect. $2$; similar constructions were recently considered in \cite{Cerejeiras}, \cite{Ablamowicz}.

The colour symmetry is somehow hidden in Nature, because the states with non-zero colour charges are not observed in experiments
due to the quark confinement mechanism (\cite{Greensite}), so that in a quark model we deal only with composite hadronic states described 
asymptotically by the Dirac and Klein-Gordon (KG) equations. We argue that on the level of single quark states one should not 
postulate the description of colour quark triplets by standard Dirac fields, and we propose instead its $12$-component colour 
generalization, incorporating the $Z_3$-grading and generating colour entanglement.

In this new approach (see Sect. $3$) free quark field components satisfy a sixth-order generalization of Klein-Gordon's equation, which factorizes into the triple product 
of the standard Klein-Gordon operator with real mass, and a pair of Klein-Gordon operators with complex-conjugated Lee-Wick masses 
(see e.g. \cite{LeeWick}, \cite{AnselmiPiva}). In such a way we obtain a set of three $Z_3$-graded mass parameters 
$m, \; jm$ and $j^2 m$, $j = e^{\frac{2 \pi i}{3}}$, covariant under the $Z_3$-symmetry acting on complex energy plane.
By a suitable choice of $12 \times 12$ colour Dirac equations one can introduce the colour-entangled quark triplet with one real mass
and a pair of complex masses $jm$ and its complex conjugate $j^2 m$ forming together a $Z_3$-graded triplet. 
\footnote{Such triplet can be also realized by the $Z_3$-graded set of generalized Wick rotations by the angles $0, 120$ and $240$ degrees.}  

Such construction permits to introduce in Sect. $4$  the vectorial realization of $Z_3$-graded Lorentz group acting on a triplet of replicas
of one and the same four-momentum vector (see also \cite{Kerner2018B}, \cite{Kerner2019B}).\footnote{In (\cite{Kerner2019B})
the $Z_3$-graded Poincar\'e algebra is introduced in an alternative way, realized on a $Z_3$-graded triplet of Minkowskian space replicas:
one real and two complex-conjugate ones.} In Sect. $5$, we present
the full description of spinorial realizations of the $Z_3$-graded Lorentz algebra which we introduced in \cite{RKJL2019} 
(see also \cite{Kerner2019B}, \cite{RKJL2020}).

In order to describe all quark symmetries, both the exact one (colour) and broken ones (flavour, generations), we consider in Sect. $6$ 
in explicit way the action of $Z_3$-graded Lorentz algebra ${\cal{L}} = {\cal{L}}^{(0)} \oplus {\cal{L}}^{(1)} \oplus {\cal{L}}^{(2)}$ 
on the sextet of generalized $12 \times 12$-component colour Dirac $\Gamma^{\mu}$ matrices which provides the irreducible representation 
of algebra ${\cal{L}}$. It appears that such a sextet requires - as a module of which ${\cal{L}}$ can act faithfully - six $12$-component 
colour Dirac fields which span $72$ quark states and takes into account all known internal symmetries of the standard quark model, describing 
colour, flavour and generations. Further, in Sect. $7$ we introduce chiral flavour 
doublets and we show how to define chiral colour spinors in the framework using the colour Dirac equations.
 
In a short outlook in final Sect. 8 we are pointing out the differences between our approach and other results
dealing with possible modifications of internal symmetries sector in the quark models. Some problems which may occur in the
procedure of supplementing our model with gauge interactions are also briefly addressed.

\section{Ternary Clifford algebra, $Z_3$-symmetry and the $SU(3)$ algebra}
\vskip 0.2cm
\indent
In a recent series of papers (\cite{RKOS2014}, \cite{AKL2017}, \cite{Kerner2018A}, \cite{Kerner2018B}, \cite{Kerner2019}) ternary
algebraic structures have been introduced and discussed. Among others, a ternary generalization of Clifford algebra with two generators 
(see e.g. in \cite{Cerejeiras}, \cite{Ablamowicz}) is of particular
interest for high energy physics due to its close relation with the Lie algebra of the $SU(3)$ group appearing as an exact colour symmetry
and as a broken symmetry mixing the three quark families.

The standard $3 \times 3$ matrix basis of ternary Clifford algebra (which was first considered in XIX-th century by Cayley \cite{Cayley} 
and Sylvester \cite{Sylvester}, who called its elements ``{\it nonions}'' ) looks as follows:
\begin{equation}
Q_1 = \begin{pmatrix}  0 & 1 & 0 \cr 0 & 0 & j \cr j^2 & 0 & 0 \end{pmatrix}, \; 
Q_2 = \begin{pmatrix}  0 & 1 & 0 \cr 0 & 0 & j^2 \cr j & 0 & 0 \end{pmatrix}, \; 
Q_3 = \begin{pmatrix}  0 & 1 & 0 \cr 0 & 0 & 1 \cr 1 & 0 & 0 \end{pmatrix}, \;  
\label{threeQ}
\end{equation} 
\begin{equation}
Q^{\dagger}_1 = \begin{pmatrix}  0 & 0 & j \cr 1 & 0 & 0 \cr 0 & j^2 & 0 \end{pmatrix}, \; 
Q^{\dagger}_2 = \begin{pmatrix}  0 & 0 & j^2 \cr 1 & 0 & 0 \cr 0 & j & 0 \end{pmatrix}, \; 
Q^{\dagger}_3 = \begin{pmatrix}  0 & 0 & 1 \cr 1 & 0 & 0 \cr 0 & 1 & 0 \end{pmatrix}, \; 
\label{threeQdagger}
\end{equation} 
where $j$ is the third primitive root of unity, 
\begin{equation} j = e^{\frac{2 \pi i}{3}}, \; \; j^2 = e^{\frac{4 \pi i}{3}}, \; \; 1+j+j^2 = 0.
\label{Jot}
\end{equation}
and ${\cal{M}}^{\dagger}$ denotes the hermitian conjugate of matrix ${\cal{M}}$. We see that all the matrices (\ref{threeQ}, \ref{threeQdagger})
are non-Hermitian.
To complete the basis of $3 \times 3$ traceless matrices, we must add to  (\ref{threeQ}) and (\ref{threeQdagger}) the following two linearly independent {\it diagonal} 
matrices:
\begin{equation}
B = \begin{pmatrix}  1 & 0 & 0 \cr 0 & j & 0 \cr 0 & 0 & j^2 \end{pmatrix}, \; \; \; \; B^{\dagger} = 
\begin{pmatrix}  1 & 0 & 0 \cr 0 & j^2 & 0 \cr 0 & 0 & j \end{pmatrix}.
\label{twoBmatrices}
\end{equation}
In what follows, we shall often use alternative notation $I_A, \; A = 1,2,...8$, with
\begin{equation}
I_1 = Q_1, \; I_2 = Q_2, I_3 = Q_3, \; I_4 = Q^{\dagger}_1, \; I_5 = Q^{\dagger}_2, \; I_6 = Q^{\dagger}_6, \; I_7 = B, \; I_8 = B^{\dagger}
\label{IAdef}
\end{equation} 
and can also add $I_0 = {\mbox{l\hspace{-0.55em}1}}_3$. The Hermitian conjugation  $I_A^{\dagger} \; \; (A = 1, 2, ..., 8):$
\begin{equation}
I_A^{\dagger} = (Q^{\dagger}_1, \; Q^{\dagger}_2, \; Q^{\dagger}_3, \; Q_1, \; Q_2, \; Q_3, \; B^{\dagger}, \; B) = I_{A^{\dagger}}
\label{Hermconjug}
\end{equation}
provides the following permutation of indices $A \rightarrow {A^{\dagger}}$:
\begin{equation}
A= (1,2,3,4,5,6,7,8) \rightarrow {A}^{\dagger} = (4,5,6,1,2,3,8,7).
\label{Apermute}
\end{equation}
We can introduce as well the standard complex conjugation ${\cal{M}} \rightarrow {\bar{\cal{M}}}$, which leads to the relations
\begin{equation}
{\bar{I}}_A = ({\bar{Q_1}} = Q_2, {\bar{Q_2}} = Q_1, {\bar{Q_3}}=Q_3, \; \; {\bar{Q}}_1^{\dagger} = Q_2^{\dagger}, \; 
 {\bar{Q}}_2^{\dagger} = Q_1^{\dagger}\; \; {\bar{B}}= B^{\dagger}) = I_{\bar{A}},
\label{QBbars}
\end{equation}
which corresponds to another permutation of indices $A$,
\begin{equation}
A= (1,2,3,4,5,6,7,8) \rightarrow {\bar{A}} = (2,1,3,5,4,6,8,7).
\label{Apermutebis}
\end{equation}

The $3 \times 3$ matrices $Q_3$ and $Q_3^{\dagger}$ are real, while $Q_2 = {\bar{Q}}_1$ are mutually complex conjugated, as well as their
Hermitean counterparts $Q_2^{\dagger} = {\bar{Q^{\dagger}_1}}$.

The matrices (\ref{threeQ}) and (\ref{threeQdagger}) are endowed with natural $\mathbb Z_3$-grading
\begin{equation}
{\rm grade} (Q_k) = 1, \; \; \; {\rm grade} (Q^{\dagger}_k) = 2,
\label{gradeQ}
\end{equation}
Out of three independent $\mathbb Z_3$-grade $0$ ternary (i.e. three-linear)
combinations, only one leads to a non-vanishing result. One can simply  check that both $j$ and $j^2$ 
{\it ternary skew commutators} do vanish
\begin{equation}
 \{ Q_1, Q_2, Q_3 \}_j = Q_1 Q_2 Q_3 + j Q_2 Q_3 Q_1 + j^2 Q_3 Q_1 Q_2 = 0, 
\end{equation}
\begin{equation}
\{ Q_1, Q_2, Q_3 \}_{j^2} = Q_1 Q_2 Q_3 + j^2 Q_2 Q_3 Q_1 + j Q_3 Q_1 Q_2 = 0,
\end{equation}
as well as the odd permutation, e.g.  $Q_2 Q_1 Q_3 + j Q_1 Q_3 Q_2 + j^2 Q_3 Q_2 Q_1 = 0$.

In contrast, the totally symmetric combination does not vanish but 
it is proportional to the  $3 \times 3$ identity matrix  $I_0 = {\mbox{l\hspace{-0.55em}1}}_3 $: 
\begin{equation}
 Q_a Q_b Q_c + Q_b Q_c Q_a + Q_c Q_a Q_b = 3\,\eta_{abc} \,  \;  {\mbox{l\hspace{-0.55em}1}}_3, \; \; \; a,b,... = 1,2,3.
\label{anticom}
\end{equation}
with $\eta_{abc}$ given by the following non-zero components
\begin{equation}
\eta_{111} = \eta_{222} = \eta_{333} = 1, \; \; \eta_{123} = \eta_{231} = \eta_{312} = j^2, \; \; 
\eta_{213} = \eta_{321} = \eta_{132} = j
\label{defeta}
\end{equation}
and all other components vanishing. 
The above relation can be used as definition of  {\it ternary Clifford algebra} (see e.g. \cite{SylvesterA}, \cite{RKOS2014}).

Analogous set of relations is formed by Hermitian conjugates $Q_{\dot{a}}^{\dagger} := {\bar{Q}}_a^T$ of matrices $Q_a$, 
which we shall endow with dotted indices ${\dot{a}}, {\dot{b}},...=1,2,3$. They satisfy the relation
\begin{equation}
Q_a^2 = Q_{\dot{a}}^{\dagger}
\label{defqbar}
\end{equation}
as well as the identities conjugate to the ones in (\ref{anticom})
 \begin{equation}
 Q^{\dagger}_{\dot{a}} Q^{\dagger}_{\dot{b}} Q^{\dagger}_{\dot{c}} + 
Q^{\dagger}_{\dot{b}} Q^{\dagger}_{\dot{c}} Q^{\dagger}_{\dot{a}} + Q^{\dagger}_{\dot{c}} Q^{\dagger}_{\dot{a}} Q^{\dagger}_{\dot{b}}
 = 3\,\eta_{{\dot{a}}{\dot{b}}{\dot{c}}} \, \; {\mbox{l\hspace{-0.55em}1}}_3,  
{\rm with} \; \;  \eta_{{\dot{a}}{\dot{b}}{\dot{c}}} = {\bar{\eta}}_{cba}.
\label{anticomdot}
\end{equation}

It is obvious that any similarity transformation of the generators $Q_a$  keeps the ternary anti-commutator (\ref{anticom})
invariant. As a matter of fact, if we define ${\tilde{Q}}_b = S^{-1} Q_b S$, with $S$ a non-singular $3 \times 3$ matrix,
 the new set of generators will satisfy the same ternary relations, because it follows that
\begin{equation}
{\tilde{Q}}_a {\tilde{Q}}_b {\tilde{Q}}_c = S^{-1} Q_a S S^{-1} Q_b S S^{-1} Q_c S = S^{-1} (Q_a Q_b Q_c) S,
\end{equation}
and on the right-hand side we have the unit matrix which commutes with all other matrices, so that 
$S^{-1} \; {\mbox{l\hspace{-0.55em}1}}_3 \; S  = {\mbox{l\hspace{-0.55em}1}}_3$.

\vskip 0.2cm
\indent
Here is the full multiplication table of the associative algebra of eight basis matrices $I_A (A=1,2,...8)$.
\vskip 0.3cm

\begin{center}
\begin{tabular}{|c|c|c|c|c|c|c|c|c|}
\hline
\raisebox{0mm}[6mm][2mm]{ \, \ \ } & $Q_1$ & $Q_2$ & $Q_3$ & $Q^{\dagger}_1$ & $ Q^{\dagger}_2$ & $Q^{\dagger}_3$ & $B$ & $B^{\dagger}$ \cr
\hline\hline
\raisebox{0mm}[6mm][2mm]{ $Q_1$ } & $ Q^{\dagger}_1$ & $j^2 Q^{\dagger}_3$ & $j Q^{\dagger}_2$ & $ {\bf 1}$ & $ B^{\dagger}$ & $B$ & $ j Q_2$ & $ j^2 Q_3$ \cr
\hline
\raisebox{0mm}[6mm][2mm]{ $Q_2$ } & $j Q^{\dagger}_3$ & $ Q^{\dagger}_2$ & $j^2 Q^{\dagger}_1$ & $ B$ & ${\bf 1}$ & $ B^{\dagger} $ & $j Q_3$ & $j^2Q_1$ \cr
\hline
\raisebox{0mm}[6mm][2mm]{ $Q_3$ } & $j^2 Q^{\dagger}_2$ & $j Q^{\dagger}_1$ & $ Q^{\dagger}_3$ & $ B^{\dagger}$ & $ B $ & ${\bf 1}$ & $ j Q_1$ & $ j^2 Q_2$ \cr
\hline
\raisebox{0mm}[6mm][2mm]{ $Q^{\dagger}_1$ } & ${\bf 1}$ & $ j^2 B $ & $ j B^{\dagger} $ & $ Q_1$ & $j^2 Q_3$ & $j Q_2$ & $ Q^{\dagger}_3$ & $ Q^{\dagger}_2$ \cr
\hline
\raisebox{0mm}[6mm][2mm]{ $Q^{\dagger}_2$ } & $j B^{\dagger} $ & ${\bf 1}$ & $ j^2 B$ & $j Q_3$ & $Q_2$ & $ j^2 Q_1$ & $ Q^{\dagger}_1$ & $ Q^{\dagger}_3$ \cr
\hline
\raisebox{0mm}[6mm][2mm]{ $Q^{\dagger}_3$ } & $j^2 B$ & $ j B^{\dagger}$ & ${\bf 1}$ & $ j^2 Q_2$ & $j Q_1$ & $Q_3$ & $Q^{\dagger}_2$ & $ Q^{\dagger}_1$ \cr
\hline
\raisebox{0mm}[6mm][2mm]{ $B$ } &$Q_2$ &$ Q_3$ & $ Q_1$ & $j Q^{\dagger}_3$ & $ j Q^{\dagger}_1$ & $ j Q^{\dagger}_2$ & $ B^{\dagger} $& ${\bf 1}$ \cr
\hline
\raisebox{0mm}[6mm][2mm]{ $ B^{\dagger} $ } & $Q_3$ & $Q_1$ & $Q_2$ & $j^2 Q^{\dagger}_2$ & $ j^2 Q^{\dagger}_3$ & $ j^2 Q^{\dagger}_1$ & ${\bf 1}$ & $ B $ \cr
\hline \hline
\end{tabular} 
\end{center} 
\vskip 0.2cm
\centerline{Table I: \,\small{The multiplication table of nonion algebra}}
\vskip 0.3cm
It is also worthwhile to note that the six matrices $Q_a$  and $Q^{\dagger}_{\dot{b}}$  together with two traceless diagonal matrices 
$B$ and $B^{\dagger}$ from (\ref{threeQ}, \ref{threeQdagger})
form the basis for certain $Z_3$-graded representation of the $\mbox{SU}(3)$-algebra, as it was shown by V. Kac in 1994 (see \cite{Kac1994}).

All these matrices are cubic roots of the $3 \times 3$ unit matrix, i.e. their cubes are all equal to  ${\mbox{l\hspace{-0.55em}1}}_3 $.
One can observe that two traceless matrices $I_2 = Q_2$ and $I_7 = B$ generate, by consecutive multiplications, full $8$-dimensional basis 
of the $SU(3)$ algebra. The full basis of $3 \times 3$ traceless $SU(3)$ matrices is generated by all possible powers and products of $B$ 
and $Q_2$, and is displayed in Table 1 below. 

We endow the two diagonal matrices $B$ and $B^{\dagger} = B^2$ with $Z_3$ grade $0$, 
the matrices $Q_a$ with $Z_3$ grade $1$, and their three hermitian conjugates ${\bar{Q}}_{\dot{b}} $ with $Z_3$ grade $2$. 
 Under matrix multiplication the grades are additive modulo $3$.

The eight matrices $B, B^{\dagger}, Q_a, Q^{\dagger}_b$ can be mapped faithfully onto the canonical Gell-Mann basis of the $SU(3)$ algebra. 
The Lie algebra of the commutators between the generators $I_A$ is given in Appendix I.
  The linear combinations of matrices $I_A$ producing the Gell-Mann matrices are given in Appendix II. 

Further we shall use the basis (\ref{IAdef}) for the description of the generators of colour algebra, which satisfies the Lie-algebraic 
relations with particular properties of complex structure constants (see Appendix I, relation {\ref{threealpha}}).
\section{The $Z_3$-graded Dirac's equation}
\vskip 0.2cm
We shall construct a generalized equation for quarks, incorporating not only
their half-integer spin and particle-antiparticle content (due to charge conjugation, producing anti-quark states),
but also the new discrete degree of freedom, the {\it colour}, taking three possible values.

Let us describe three different two-component fields (Pauli spinors), which will be distinguished by three colours, 
the ``red" for  $\varphi_{+}$, the  ``blue" for $\chi_{+}$, and the ``green" for  $\psi_{+}$; more explicitly
\begin{equation}
 \varphi_{+} = \begin{pmatrix} \varphi_{+}^1 \cr \varphi_{+}^2 \end{pmatrix}, \; 
\chi_{+} = \begin{pmatrix} \chi_{+}^1 \cr \chi_{+}^2 \end{pmatrix}, \;   
 \psi_{+} = \begin{pmatrix} \psi_{+}^1 \cr \psi_{+}^2 \end{pmatrix}. 
\label{phichipsi1}
\end{equation}
We follow the minimal scheme which takes into account the existence of spin by using Pauli spinors on which 
the $3$-dimensional momentum operator acts through $2 \times 2$ matrix describing the scalar product ${\boldsymbol{\sigma}}\cdot {\bf p}$. 

To acknowledge the existence of anti-particles, we should also introduce three
``anti-colours", denoted by a ``minus" underscript, corresponding to
``cyan" for $\varphi_{-}$, ``yellow" for $\chi_{-}$ and ``magenta" for $\psi_{-}$; here, too,
we employ the two-component columns: 
\begin{equation}
\varphi_{-} = \begin{pmatrix} \varphi_{-}^1 \cr \varphi_{-}^2 \end{pmatrix},  \; \;  
\chi_{-} = \begin{pmatrix} \chi_{-}^1 \cr \chi_{-}^2 \end{pmatrix}, \; \;  
\psi_{-} = \begin{pmatrix} \psi_{-}^1 \cr \psi_{-}^2 \end{pmatrix}.
\label{phichipsi2}
 \end{equation}
As a result, the six Pauli spinors (\ref{phichipsi1}) and (\ref{phichipsi2}) will form a {\it twelve}-component entity which we shall call
``coloured Dirac spinor''.
This construction reflects the overall $Z_3 \times Z_2 \times Z_2$ symmetry: one  $Z_2$ group corresponds to the spin $\frac{1}{2}$ 
dichotomic degree of freedom, described by eigenstates; the second $Z_2$ is required in  order to represent the 
particle-anti-particle symmetry, and the $Z_3$ group corresponding to colour symmetry.

The ``coloured" Pauli spinors  should satisfy first order equations conceived 
in such a way that they propagate all together as one geometric object, just like  ${\bf E}$  
and ${\bf B}$ components of Maxwell's tensor in electrodynamics, or the pair of two-component Pauli spinors 
which are not propagating separately, but constitute one single entity, the four-component Dirac spinor.  

This leaves not much space for the choice of the system of intertwined equations. Here we present
the ternary generalization of Dirac's equation, intertwining not only particles with antiparticles,
but also the three ``colours" in such a way that the entire system becomes invariant under
the action of the $Z_3 \times Z_2 $ group. 

The set of linear equations for three Pauli spinors endowed with colours, and another three Pauli spinors
corresponding to their anti-particles characterized by "anti-colours" involves together twelve complex functions. 
The twelve components could describe three independent Dirac particles, but here they are intertwined in
a particular $Z_3 \times Z_2 $ graded manner, mixing together not only particle-antiparticle states, 
but the three colours as well.

Let us follow the logic that led from Pauli's to ($Z_2$-graded) Dirac's equation and extend it to the colours acted upon
by the $Z_3$-group. In the expression for the energy operator (Hamiltonian) the mass term is positive when
it describes particles, and acquires negative sign when we pass to anti-particles, i.e. one gets the change of sign 
each time when particle-antiparticle components are interchanged. 

We shall now assume that mass terms should acquire the factor $j$ when we switch
from the red component $\varphi$ to the blue component $\chi$, and 
another $j$-factor when we switch from blue component $\chi$ to the green component $\psi$.
We remind that we use the notation introduced in (\ref{Jot}), 
$ j = e^{\frac{2 \pi i}{3}}, \; \; j^2 = e^{\frac{4 \pi i}{3}}, \; \; j^3  = 1, \; \; \; {\rm and} \; \; 1+ j + j^2 = 0.$

The momentum operator will be non-diagonal, as in the Dirac equation, systematically intertwining not only
particles with antiparticles, but also colours with anti-colours. The system that satisfies all these assumptions can be introduced in the
following manner (\cite{RKOS2014}, \cite{Kerner2018B}): 

Let us first choose the basis in which particles with a given colour and the particles with corresponding anti-colour are grouped in pairs:
\begin{equation}
\left( \varphi_{+}, \varphi_{-}, \chi_{+}, \chi_{-}, \psi_{+}, \psi_{-} \right)^T. 
\label{basis1}
\end{equation}
where $ \varphi_{\pm}, \;\chi_{\pm}$ and $\psi_{\pm}$ are two-component Pauli spinors defined by eqs. (\ref{phichipsi1})
and (\ref{phichipsi2}).

In such a basis our ``coloured Dirac equation" takes the following form in terms of six Pauli spinors:
\begin{equation}
\begin{split}
& E \; \varphi_{+} = mc^2 \, \varphi_{+} + c \; {\boldsymbol{\sigma}} \cdot {\bf p} \, \chi_{-},
\\
&E \; \varphi_{-} = - mc^2 \, \varphi_{-} + c \; {\boldsymbol{\sigma}} \cdot {\bf p} \, \chi_{+}
\\
&E \; \chi_{+} = j \; mc^2 \, \chi_{+} + c \; {\boldsymbol{\sigma}} \cdot {\bf p} \, \psi_{-},
\\
&E \; \chi_{-} = - j \; mc^2 \, \chi_{-} + c \; {\boldsymbol{\sigma}} \cdot {\bf p} \, \psi_{+}
\\
&E \; \psi_{+} = j^2 \;  mc^2 \, \psi_{+} + c \; {\boldsymbol{\sigma}} \cdot {\bf p} \, \varphi_{-},
%\begin{equation}
\\
&E \; \psi_{-} = -j^2 \;mc^2 \, \psi_{-} + c \; {\boldsymbol{\sigma}} \cdot {\bf p} \, . \varphi_{+}
\end{split}
\label{systemsix}
\end{equation}
Let us remark that while in the Schroedinger picture the energy $E$ and the momentum ${\bf p}$ are represented by differential
operators
\begin{equation}
E \rightarrow - i \hbar \frac{\partial}{\partial t}, \; \; \; {\bf p} \rightarrow - i \hbar {\boldsymbol{\nabla}},
\label{Epdef}
\end{equation} 
in (\ref{systemsix}) we use their Fourier-transformed image, in which $E$ and ${\bf p}$ are interpreted as multiplication by the corresponding
numerical eigenvalues.

The particle-antiparticle $Z_2$-symmetry is obtained if $m \rightarrow -m$ and simultaneously 
$(\varphi_{+}, \chi_{+}, \psi_{+} ) \rightarrow (\varphi_{-}, \chi_{-}, \psi_{-})$
 and vice versa; the $Z_3$-colour symmetry is realized by multiplication of mass $m$ by $j$ each time the colour changes,
i.e. more explicitly, $Z_3$ symmetry is realized by the following mappings:
\begin{equation}
m \rightarrow jm, \; \; \; \varphi_{\pm} \rightarrow \chi_{\pm} \rightarrow \psi_{\pm} \rightarrow \varphi_{\pm},
\label{Z3first}
\end{equation}
\begin{equation}
m \rightarrow j^2 m, \; \; \; \varphi_{\pm} \rightarrow \psi_{\pm} \rightarrow \chi_{\pm} \rightarrow \varphi_{\pm}.
\label{Z3second}
\end{equation}
The system of equations (\ref{systemsix}) can be written using $12 \times 12$ matrices acting on 
the $12$-component colour spinor $\Psi$ build up from six ``coloured"  Pauli spinors. In shortened form we can write
\begin{equation}
{\cal E} \;  \Psi = \left[ c^2 \; {\cal{M}} + c {\cal{P}} \right] \; \Psi,
\label{EMPshort}
\end{equation}
where ${\cal{E}} = E \; {\mbox{l\hspace{-0.55em}1}}_{1 2}$, with \; \; ${\mbox{l\hspace{-0.55em}1}}_{12}$ denoting the $12 \times 12$ unit matrix, 
and the matrices ${\cal{M}}$ and ${\cal{P}}$ given explicitly below:
{\small 
\begin{equation}
{\cal{M}} =  \begin{pmatrix} m \; {\mbox{l\hspace{-0.55em}1}}_{2} & 0 & 0 & 0 & 0 & 0 \cr 0 & - m \; {\mbox{l\hspace{-0.55em}1}}_{2} & 0 & 0 & 0 & 0  \cr
0 & 0 & j m \; {\mbox{l\hspace{-0.55em}1}}_{2} & 0 & 0 & 0 \cr 0 & 0 & 0 & - j m \; {\mbox{l\hspace{-0.55em}1}}_{2} & 0 & 0 \cr 
0 & 0 & 0 & 0 & j^2 m \; {\mbox{l\hspace{-0.55em}1}}_{2} & 0 \cr
0 & 0 & 0 & 0 & 0 & -j^2 m \; {\mbox{l\hspace{-0.55em}1}}_{2} \end{pmatrix} 
\label{MMatrix}
\end{equation}
\begin{equation} 
{\cal{P}} = \begin{pmatrix} 0 & 0  & 0 & {\boldsymbol{\sigma}} \cdot {\bf p} & 0 & 0   \cr 
0 & 0 &  {\boldsymbol{\sigma}} \cdot {\bf p} & 0 & 0 & 0 \cr
0 & 0 & 0 & 0 & 0 & {\boldsymbol{\sigma}} \cdot {\bf p}  \cr 
0 & 0 & 0 & 0 & {\boldsymbol{\sigma}} \cdot {\bf p} & 0 \cr 
0 & {\boldsymbol{\sigma}} \cdot {\bf p} & 0 & 0 & 0 & 0 \cr
{\boldsymbol{\sigma}} \cdot {\bf p} & 0 & 0 & 0 & 0 & 0 \end{pmatrix}
\label{PMatrix}
\end{equation}}

The two matrices ${\cal{M}}$ and ${\cal{P}}$ in (\ref{MMatrix}) and (\ref{PMatrix}) are $12 \times 12$-dimensional: all the entries in ${\cal{M}}$
are proportional to the $2 \times 2$ unit matrix, 
and the entries in the second matrix ${\cal{P}}$ contain $2 \times 2$ Pauli's sigma-matrices, so ${\cal{P}}$ is as well a $12 \times 12$ matrix. 
The energy matrix operator ${\cal{E}}$ is proportional to the  $12 \times 12$ unit matrix. 

One can easily see that the diagonalization of the system is achieved only at the sixth iteration.
The final result is extremely simple: all the components satisfy the same sixth-order equation, 

$$ E^6 \; \varphi_{+} = m^6 c^{12} \; \varphi_{+} + c^6 \mid {\bf p} \mid^6 \; \varphi_{+}, $$
\begin{equation}
E^6 \; \varphi_{-} = m^6 c^{12} \; \varphi_{-} + c^6 \mid {\bf p} \mid^6 \; \varphi_{-}.
\label{E6varphi}
\end{equation}
and similarly all other components.
It is convenient to use the tensor product notation for the description of the matrices ${\cal{E}}, \; {\cal{M}}$ and ${\cal{P}}$.

Using two $3 \times 3$ matrices $B$ and $Q_3$ defined in (\ref{threeQ}),
\begin{equation}
B = \begin{pmatrix} 1 & 0 & 0 \cr 0 & j & 0 \cr 0 & 0 & j^2 \end{pmatrix} \; \; {\rm and} \; \; 
Q_3 = \begin{pmatrix} 0 & 1 & 0 \cr 0 & 0 & 1 \cr 1 & 0 & 0 \end{pmatrix},
\label{BQmatrices}
\end{equation} 
the $12 \times 12$ matrices $M$ and $P$ can be represented as the following  tensor products:
\begin{equation}
{\cal{E}} = E \; {\mbox{l\hspace{-0.55em}1}}_3 \otimes {\mbox{l\hspace{-0.55em}1}}_2 \otimes {\mbox{l\hspace{-0.55em}1}}_2
 = E \; {\mbox{l\hspace{-0.55em}1}}_{12}, \; \; \; \; 
{\cal{M}} = m \; B \otimes \sigma_3 \otimes {\mbox{l\hspace{-0.55em}1}}_2, \; \; \; \; \; \;  
{\cal{P}} =  Q_3 \otimes \sigma_1 \otimes  ({\boldsymbol{\sigma}} \cdot {\bf p}) 
\label{MPtensor1}
\end{equation}
where \; \; $ {\mbox{l\hspace{-0.55em}1}}_2 = \begin{pmatrix} 1 & 0 \cr 0 & 1 \end{pmatrix}, \; \; \; 
\sigma_1 = \begin{pmatrix} 0 & 1 \cr 1 & 0 \end{pmatrix}, \; \; \; 
\sigma_3 = \begin{pmatrix} 1 & 0 \cr 0 & -1 \end{pmatrix}. $  

Let us rewrite the system (\ref{systemsix}) involving six coupled two-component spinors as one linear equation for the
``colour Dirac spinor'' $\Psi$, conceived as  column vector containing twelve components of three ``colour" fields, in the basis (\ref{basis1})
given by$\Psi = [\varphi_{+}, \varphi_{-}, \chi_{+}, \chi_{-}, \psi_{+}, \psi_{-} ]^T$,
with energy and momentum operators ${\cal{E}}$ and ${\cal{P}}$ on the left hand side and the mass operator ${\cal{M}}$ on the right hand side:
\begin{equation}
E \; {\mbox{l\hspace{-0.55em}1}}_3 \otimes {\mbox{l\hspace{-0.55em}1}}_2 \otimes {\mbox{l\hspace{-0.55em}1}}_2 \; \Psi 
-  Q_3 \otimes \sigma_1 \otimes c \; {\boldsymbol{\sigma}}\cdot {\bf p} \; \Psi
= m c^2 \; B \otimes \sigma_3 \otimes {\mbox{l\hspace{-0.55em}1}}_2 \; \Psi.
\label{EPtogether}
\end{equation}
Like in the case of the standard Dirac equation, let us transform this equation
in a way that the mass operator becomes proportional the the unit matrix. For such a purpose, we multiply the equation
(\ref{EPtogether}) on the left by the matrix $ B^{\dagger} \otimes \sigma_3  \otimes {\mbox{l\hspace{-0.55em}1}}_2.$

Now we get the following equation which enables us to interpret the energy and the momentum as components of 
the Minkowskian four-vector $c \; p^{\mu} = [E, \; c {\bf p}] $: 
\begin{equation}
E \;  B^{\dagger} \otimes \sigma_3 \otimes {\mbox{l\hspace{-0.55em}1}}_2 \; \Psi
- Q_2 \otimes (i \sigma_2) \otimes c \, {\boldsymbol{\sigma}}\cdot {\bf p} \; \Psi= 
m c^2 \;  {\mbox{l\hspace{-0.55em}1}}_3 \otimes {\mbox{l\hspace{-0.55em}1}}_2 \otimes {\mbox{l\hspace{-0.55em}1}}_2 \; \Psi,
\label{Gammafirst}
\end{equation}
where we used the fact that 
$(\sigma_3)^2  = {\mbox{l\hspace{-0.55em}1}}_2$,\;
$B^{\dagger} B = {\mbox{l\hspace{-0.55em}1}}_3$ and $B^{\dagger} Q_3 =  Q_2$.
The sixth power of this operator gives the same result as before,
\begin{equation}
\left[ E \;  B^{\dagger} \otimes \sigma_3 \otimes {\mbox{l\hspace{-0.55em}1}}_2
- Q_2 \otimes (i \sigma_2) \otimes c \, {\boldsymbol{\sigma}}\cdot {\bf p} \right]^6 =
\left[ E^6 - c^6 {\bf p}^6 \right] \; {\mbox{l\hspace{-0.55em}1}}_{12} = m^6 c^{12} \; {\mbox{l\hspace{-0.55em}1}}_{12}
\label{sixpower1}
\end{equation}
It is also worth to note that taking the determinant on both sides of the eq. (\ref{Gammafirst}) yields the twelfth-order equation:
\begin{equation} 
\left( E^6 - c^6 \; \mid {\bf p} \mid^6 \right)^2 = m^{12} c^{24}.
\label{dettwelve}
\end{equation}

There is still certain arbitrariness in the choice of $3 \times 3$ matrix factors $B^{\dagger}$ and $Q_2$ in the colour Dirac operator 
(\ref{Gammafirst}). This is due to the choice of $j = e^{\frac{2 \pi i}{3}}$ as the generator of the representation of the finite $Z_3$-symmetry group. 
If $j^2$ is chosen instead, in (\ref{Gammafirst}) the matrix $B^{\dagger}$ will be replaced by $B$, $Q_2$ by $Q_1$, which is
its complex conjugate; the remaining terms keep the same form.  

The equation (\ref{Gammafirst}) can be written in a concise manner by introducing the $12 \times 12$ matrix colour Dirac operator
$\Gamma^{\mu} p_{\mu}$ using Minkowskian space-time indices and metric $\eta_{\mu \nu} = {\rm diag} (+, -, -, - )$:
\begin{equation}
\Gamma^{\mu} p_{\mu} \; \Psi = m c \; \; {\mbox{l\hspace{-0.55em}1}}_{12} \; \Psi, \; \; \; 
{\rm with} \; \; p^0 = \frac{E}{c}, \; \; p^k = [ \; p^x, p^y, p^z \; ].
\label{Gammasecond}
\end{equation}
with $12 \times 12$ matrices $\Gamma^{\mu} \; \; (\mu = 0, 1, 2, 3)$ defined as follows:
\begin{equation}
\Gamma^0 =  B^{\dagger} \otimes \sigma_3 \otimes {\mbox{l\hspace{-0.55em}1}}_2, \; \; \; \; \; 
\Gamma^{k} =  Q_2 \otimes (i \sigma_2) \otimes  {\sigma}^k
\label{Gammasbig}
\end{equation}
The multiplication rules for $B$, $B^{\dagger}$, $Q_A$ and $Q_B^{\dagger}$, \; ($A,B,... = 1,2,3$) are given in the Table 1.

The $12$-component colour Dirac equation (\ref{Gammasecond}) is invariant under an arbitrary similarity transformation, i.e. if we set
\begin{equation}
\Psi' = {\cal{R}} \; \Psi, \; \; \; (\Gamma^{\mu})' = {\cal{R}} \; \Gamma^{\mu} \; {\cal{R}}^{-1} \; \; \; 
{\rm then} \; \; \; (\Gamma^{\mu})' p_{\mu} \; \Psi' = mc \; \Psi',
\label{similgamma}
\end{equation}
we get obviously 
\begin{equation}
\left[ (\Gamma^{\mu})' p_{\mu} \right]^6 = (p_0^6 - \mid {\bf p} \mid^6 ) \; {\mbox{l\hspace{-0.55em}1}}_{12}
\label{Diracprim}
\end{equation}
Following the formulae (\ref{Gammasbig}) for the colour Dirac $\Gamma^{\mu}$-matrices we see that they are neither real 
(${\bar{\Gamma}}^{\mu} \neq \Gamma^{\mu}$) nor Hermitian ($(\Gamma^{\mu})^{\dagger} \neq \Gamma^{\mu}$). From the colour
Dirac equation (\ref{Gammafirst}) one gets the following equations for complex-conjugated ${\bar{\Psi}}$ and Hermitean-conjugated $\Psi^{\dagger}$:
\begin{equation}
{\bar{\Gamma}}^{\mu} p_{\mu} \; {\bar{\Psi}} = mc \; {\bar{\Psi}}, \; \; \; \; \; \; 
p_{\mu} \Psi^{\dagger} (\Gamma^{\mu})^{\dagger} = mc \Psi^{\dagger},
\label{PsiPsibar}
\end{equation}
where ${\bar{\Psi}}$ is a column, $\Psi^{\dagger}$ is a row, ${\bar{\sigma}}_k = - \sigma_2 \sigma_k \sigma_2$, $\sigma_k = \sigma^k, \; \sigma_0 = \sigma^0 = 
{\mbox{l\hspace{-0.55em}1}}_{2}$, and
$${\bar{\Gamma}}^0 = B \otimes \sigma_3 \otimes \; {\mbox{l\hspace{-0.55em}1}}_{2}, \; \; \; 
{\bar{\Gamma}}^k = Q_1 \otimes (i \sigma_2) \otimes {\bar{\sigma}}^k,$$
\begin{equation}
(\Gamma^0)^{\dagger} = B \otimes \sigma_3 \otimes \;  {\mbox{l\hspace{-0.55em}1}}_{2}, \; \; \;
(\Gamma^k)^{\dagger} = Q_1 \otimes \sigma_3 \otimes \sigma^k,
\label{Gammasbar}
\end{equation}
Further, the second equation of (\ref{PsiPsibar}) can be written in terms of the matrices (\ref{Gammasbig}) if we introduce the Hermitian-adjoint 
colour Dirac spinor $\Psi^H = \Psi^{\dagger} C$, where the $12 \times 12$-matrix $C$ satisfies the relation
\begin{equation}
(\Gamma^{\mu})^{\dagger} C = C \Gamma^{\mu}.
\label{CGammaC}
\end{equation}
It can be also shown that neither ${\bar{\Gamma}}^{\mu}$ nor $(\Gamma^{\mu})^{\dagger}$
can be obtained via similarity transformation (\ref{similgamma}).

\begin{figure}[hbt]
\centering 
\includegraphics[width=6.6cm, height=5.2cm]{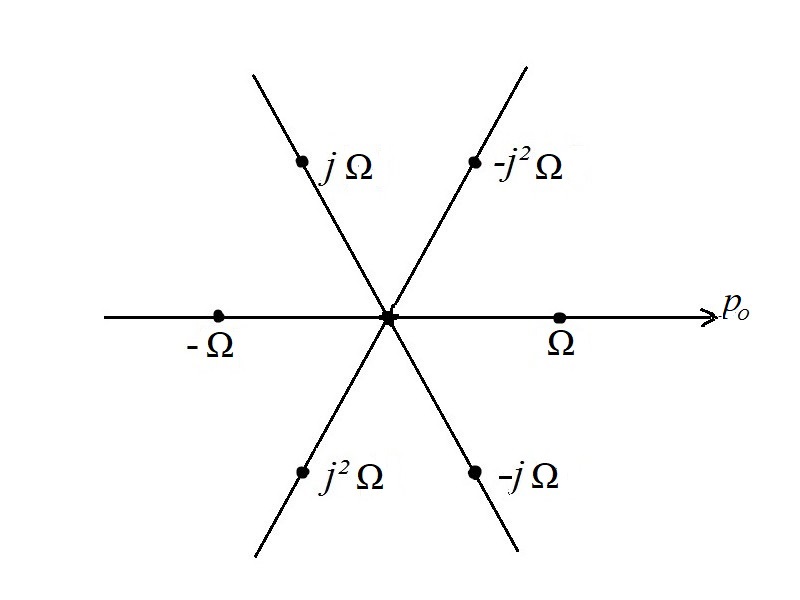}
\caption{{\small The six simple poles in the Fourier-transform of the propagator (\ref{InverseK}), with two real ones $\pm \Omega$ and two 
conjugate Lee-Wick poles $\pm j \Omega, \; \pm j^2 \Omega$. The Lee-Wick complex-conjugate masses were introduced in \cite{LeeWick}; 
see also \cite{AnselmiPiva}.}}
\label{fig:Sixpoints}
\end{figure}

To obtain a general solution of the colour Dirac equation one should use its Fourier transformed version (see (\ref{Gammasecond})). 
In the momentum space it becomes:
\begin{equation}
\left( \Gamma^{\mu} \, p_{\mu} - m \; {\mbox{l\hspace{-0.55em}1}}_{12} \right) \; {\hat{\Psi}} (p) = 0.
\label{Dirterfour}
\end{equation}
The sixth power of the matrix $\Gamma^{\mu} p_{\mu}$ is diagonal and proportional to $m^6$, so that we have
\begin{equation}
 \left( \Gamma^{\mu} p_{\mu} \right)^6 - m^6  \; {\mbox{l\hspace{-0.55em}1}}_{12}   =
\left( p_0^6 - \mid {\bf p} \mid^6 - m^6 \right) \;  {\mbox{l\hspace{-0.55em}1}}_{12} =  0.
\label{Gammasixem}
\end{equation}
Now we should find the inverse of the matrix  $\left( \Gamma^{\mu} \, p_{\mu} - m \; {\mbox{l\hspace{-0.55em}1}}_{12} \right)$.
Let us note that the sixth-order expression on the left-hand side in (\ref{Gammasixem}) can be factorized as follows:
\begin{equation}
\left( \Gamma^{\mu} p_{\mu} \right)^6 - m^6 = \left( \left(\Gamma^{\mu} p_{\mu} \right)^2 - m^2 \right) \;
\left(  \left( \Gamma^{\mu} p_{\mu} \right)^2 - j \; m^2 \right) \; \left( \left( \Gamma^{\mu} p_{\mu} \right)^2 - j^2 \; m^2 \right).
\label{Gammafact3}
\end{equation}
The first factor can be expressed as the product of two linear operators, one of which defines the colour Dirac equation (\ref{Gammasecond})
(see also (\ref{Dirterfour}):
\begin{equation}
\left( \Gamma^{\mu} p_{\mu} \right)^2 - m^2 =
\left( \Gamma^{\mu} p_{\mu} - m \right) \;\left( \Gamma^{\mu} p_{\mu} + m \right) \;
\label{Gammafact4}
\end{equation}
Therefore the inverse of the Fourier transform of the linear operator defining the colour Dirac equation (\ref{Dirterfour}) is given 
by the following matrix:
\begin{equation}
\left[ \Gamma^{\mu} p_{\mu}  - m \right]^{-1} =
\frac{\left( \Gamma^{\mu} p_{\mu} + m \right) \;
\left(  \left( \Gamma^{\mu} p_{\mu} \right)^2 - j \; m^2 \right) \;
\left( \left( \Gamma^{\mu} p_{\mu} \right)^2 - j^2 \; m^2 \right)}{\left( p_0^6 - \mid {\bf p} \mid^6 - m^6 \right)}.
\label{Dirac3inverse}
\end{equation} 
The inverse of the six-order polynomial can be decomposed into a sum of three expressions with second-order denominators, 
multiplied by the common factor of the fourth order. Let us denote by $\Omega$ the sixth root of $( \mid {\bf p} \mid^6 + m^6 )$,
\begin{equation}
\Omega = \root6\of{ \mid {\bf p} \mid^6 + m^6},
\label{BigOmega}
\end{equation}
along with five other root values obtained via multiplication by consecutive powers of the sixth root of unity, $q = e^{\frac{2 \pi i}{6}}$.
Recalling relation (\ref{Jot}) and that $q^2 = j$, we have the identity
\begin{equation}
(p_0^6 - \Omega^6) = (p_0^2 - \Omega^2)((p_0^2 - j \Omega^2)((p_0^2 - j^2 \Omega^2) 
%= (p_0^2 - \Omega^2)((j^2 p_0^2 - \Omega^2)((j p_0^2 - \Omega^2),
\label{Decomp1}
\end{equation}
which leads to the decomposition formula
\begin{equation}
\frac{1}{\left( p_0^6 - \mid {\bf p} \mid^6 - m^6 \right)} = \frac{1}{3 \; \Omega^4} \left[ \frac{1}{p_0^2 - \Omega^2} +
+ \frac{j}{ p_0^2 - j \Omega^2} + \frac{j^2}{ p_0^2 - j^2 \Omega^2} \right]
\label{InverseK}
\end{equation}
or equivalently,
\begin{equation}
\frac{1}{\left( p_0^6 - \mid {\bf p} \mid^6 - m^6 \right)} = \frac{1}{3 \; \Omega^4} \left[ \frac{1}{p_0^2 - \Omega^2} +
+ \frac{1}{j^2 p_0^2 - \Omega^2} + \frac{1}{j p_0^2 - \Omega^2} \right]
\label{InverseK2}
\end{equation}
As long as there is a non-zero mass term, we do not encounter the infrared divergence problem at $\mid {\bf p} \mid \rightarrow 0$.
Each of the three inverses of a second-order polynomial can be in turn expressed as a sum of simple first-order poles, e.g. 
\begin{equation}
\frac{1}{p_0^2 - j \Omega^2} = \frac{j}{2 \; \Omega} \left[ \frac{1}{p_0 -j^2 \Omega} - \frac{1}{p_0 + j^2 \Omega} \right]
= \frac{j^2}{2 \; \Omega} \left[ \frac{1}{j p_0 - \Omega} - \frac{1}{j p_0 + \Omega} \right],
\label{decompK}
\end{equation}
and similarly for other terms in (\ref{InverseK}). After such a substitution in (\ref{Dirac3inverse}), six $Z_6$-graded simple poles do appear,
 Figure (\ref{fig:Sixpoints}) illustrating the location of these six poles in the complex energy plane. 
In order to introduce the propagators in the coordinate space, one has to perform the contour integrals in complex energy plane. 

The first term in the decomposition (\ref{InverseK}) of the coulour Dirac propagator (\ref{InverseK}) presents two simple poles on the real line, 
while the second and the third terms display two simple poles each, located on complex straight lines $Im p_0 = j Re p_0$ and $Im p_0 = j^2 Re p_0$.

One can add that in the propagators given by formula (\ref{InverseK}) the non-standard residua $\pm j$ and $\pm j^2$ should be justified
by suitable form of the $Z_3$-graded commutators describing quantum oscillator algebra of colour quark field excitations. 

It should be stressed that the colour Dirac equation (\ref{PsiPsibar}) breaks the Lorentz symmetry ${\cal{O}} (1,3)$ reducing it 
to ${\cal{O}}_3$, because the $3 \times 3$-matrices describing  ``colour'' are {\it different} 
for the $\Gamma^0$ and $\Gamma^k$ components. 
However we shall show in the following Section $4$ that one can introduce a $Z_3$-graded generalization of the Lorentz transformations, 
acting in covariant way on three ``replicas'' of the energy-momentum four-vector introduced above. Analogous extensions of space-time were 
discussed in \cite{Finkelstein}, \cite{MarschNarita}.

\section{$Z_3$-graded set of three complex four-momenta and $Z_3$-graded Lorentz transformations}
\vskip 0.2cm
\indent
The mass shell condition (\ref{dettwelve}) for coloured Dirac equation can be decomposed into the usual relativistic Klein-Gordon invariant
multiplied by a strictly positive factor which can be interpreted as generating the form-factor for quark propagator with given mass $m$.
\begin{equation}
C_6 = p_0^6 - \Omega^6 = (p_0^2 - \mid {\bf p} \mid^2)(p_0^4 + p_0^2 \mid {\bf p} \mid^2 + \mid {\bf p} \mid^4) = m^6 c^6,
\label{C6withform}
\end{equation}
The sixth-order polynomian $C_6$ can be further decomposed into the product of the following three second-order polynomials, 
\begin{equation}
C_6 = {\overset{(0) \; \; \; }{C_{2}}}{\overset{(1) \; \; \; }{C_{2}}}{\overset{(2) \; \; \; }{C_{2}}}, 
\label{Csix}
\end{equation}
with
\begin{equation}
{\overset{ \; (0) \; \; \; }{C_{2}}} = p_0^2 - {\bf p}^2, \; \; \; {\overset{ \; (1) \; \; \; }{C_{2}}} = j \;  p_0^2 - {\bf p}^2, \; \; 
{\overset{ \; (2) \; \; \; }{C_{2}}} = j^2 \; p_0^2 - {\bf p}^2.
\label{threeC2}
\end{equation}
Let us denote by superscripts $(0), \; (1)$ and $(2)$ the four-momenta with quadratic invariants given by
$ {\overset{ \; (0) \; \; \; }{C_{2}}}, \; {\overset{ \; (1) \; \; \; }{C_{2}}}$ and ${\overset{ \; (2) \; \; \; }{C_{2}}}$.

We get explicitly
\begin{equation}
(p_0)^2 - ({\overset{(0)}{\bf p}})^2 = {\overset{ \; (0) \; \; \; }{C_{2}}}, \; \; \; 
({\overset{(1)}{p_{0}}})^2 - ({\overset{(1)}{\bf p}})^2 = {\overset{ \; (1) \; \; \; }{C_{2}}}, \; \; \; 
({\overset{(2)}{p_{0}}})^2 - ({\overset{(2)}{\bf p}})^2 = {\overset{ \; (2) \; \; \; }{C_{2}}}, \; \; \; 
\label{threepisinv}
\end{equation}
From any real four-vector ${\overset{(0) }{p_{0}}}_{\mu}$ one can define  its two ``replicas'' $0{\overset{(1)}{\bf p}}_{\mu}$ and 
${\overset{(2)}{\bf p}}_{\mu}$ with $p_0$ in the complex plane, obtained by the generalized 
Wick rotations by $j$ and by $j^2$ in the following way. Let us introduce three $4 \times 4$ matrices acting on Minkowskian four-vectors:
\begin{equation}
{\overset{(0) \; \; }{A}} = {\rm diag} \; (1,1,1,1) = {\mbox{l\hspace{-0.55em}1}}_{4} , \; \;  
{\overset{(1) \; \; }{A}} = {\rm diag} \; (j^2,1,1,1), \; \; 
{\overset{(2) \; \; }{A}} = {\rm diag} \; (j,1,1,1),
\label{ThreeA}
\end{equation}
providing a (reducible) matrix representation of the cyclic $Z_3$ group,
\begin{equation}
{\overset{(r)}{A}} {\overset{(s)}{A}} = {\overset{(r+s)}{A}}.
\label{AZ3}
\end{equation}
where the superscripts $(r+s)$ are added modulo $3$, e.g. $1 + 2 \rightarrow  0, \; 2 + 2 \rightarrow 1$, etc.
Acting on a given four-vector $p_{\mu} = (p_0, {\bf p})$ by one of the matrices ${\overset{(r) }{A}}$ we produce its three $Z_3$-graded ``replicas''
belonging correspondingly to sectors ${\overset{ \; (r) \; \; \; }{C_{2}}}$:
%symbolically, ${\overset{(r) }{A}} p = {\overset{(r) }{p}}$
\begin{equation}
{\overset{(r) }{p}} = {\overset{(r) }{A}} p: \; \; \; 
{\overset{(0) \; \; }{p_{\mu}}} \rightarrow \begin{pmatrix} p_0 \cr {\bf p} \end{pmatrix}, \; \; 
{\overset{(1) \; \; }{p_{\mu}}} \rightarrow \begin{pmatrix} j^2 p_0 \cr {\bf p} \end{pmatrix}, \; \; 
{\overset{(2) \; \; }{p_{\mu}}} \rightarrow \begin{pmatrix} j p_0 \cr {\bf p} \end{pmatrix}.
\label{gradedpis}
\end{equation}
In what follows, we shall use for both the Lorentz boosts and the Wick rotations a short-hand notation:
\begin{equation}
{p'}_{\mu}  = L_{\mu}^{\; \; \nu} p_{\nu}  \rightarrow p' = L p, \; \; \; \; \; \; \; {\overset{(r) }{{p'}_{\mu}}} = {\overset{(r) }{{A_{\mu}^{\; \; \nu}}}} 
{\overset{(0) }{p_{\nu}}} \rightarrow {\overset{(r) }{p'}} = {\overset{(r) }{A}} {\overset{(0) }{p}} 
\label{SymbolicLA}
\end{equation}
It should be stressed here that the spacetime remains Minkowskian, with one real time and three real spatial coordinates; however, the
components of  ${\overset{(1) \; \; }{p_{\mu}}}$ and ${\overset{(2) \; \; }{p_{\mu}}}$
can take on particular $Z_3$-graded complex values. Three ``replicas'' (\ref{gradedpis}) are the images of the same four-vector
which can be obtained by $Z_3$-valued Wick rotations in complex energy plane.
 
In particular, let us denote by ${\overset{(0) \; \; \; }{L_{00}}}$ the classical Lorentz transformations which map the real Minkowskian momenta 
${\overset{(0) \; \; }{p_{\nu}}}$ into ${\overset{(0) \; \; }{p^{'}_{\nu}}}$

\begin{equation}
( {\overset{(0) \; \; \; }{L_{00}}} )_{\mu}^{\; \; \nu} {\overset{(0) \; \; }{p_{\nu}}} = {\overset{(0) \; \; }{{p'}_{\mu}}} 
\rightarrow {\overset{(0) \; \; \; }{L_{00}}} {\overset{(0) \; \; }{p}} = {\overset{(0) \; \; }{p'}},
\label{Lzeroshort}
\end{equation}
where lower indices $(00)$ mean that we transform ${\overset{(0) \; \; \; }{C_{2}}}$ into itself, and the superscript
$(0)$ says that we deal with the classical Lorentz transformations.
The zero-grade Lorentz transformations can be extended to the mappings of four-vectors ${\overset{(r) \; \; }{p}}$ belonging to sector                                                                                    
${\overset{(r) \; \; \; }{C_{2}}}$ onto four-vectors  ${\overset{(s) \; \; }{p'}}$ belonging to sector ${\overset{(s) \; \; \; }{C_{2}}}$, with $r, s = 0, 1, 2$. 
Let us apply a Lorentz boost transforming a four-vector
from the sector $s$, $ {\overset{(s) \; \; }{p}}$ into a vector from another sector $r$,  ${\overset{(r) \; \; }{p'}}$. Using the shorthand  
notations (\ref{SymbolicLA}) and (\ref{Lzeroshort}), we have:
\begin{equation}
{\overset{(r) \; \; }{p'}} =  {\overset{(r) \; \; }{A}} {\overset{(0) \; \; }{p}} =  
{\overset{(r) \; \; }{A}} {\overset{(0) \; \; \; }{L_{00}}}  {\overset{(0) \; \; }{p}} =
{\overset{(r) \; \; }{A}} {\overset{(0) \; \; \; }{L_{00}}} {\overset{(s) \; \; }{A^{-1}}} {\overset{(s) \; \; }{A}}  {\overset{(0) \; \; }{p}}
= {\overset{(r-s) \; \; \; }{L_{rs}}} {\overset{(s) \; \; }{p}}
\label{Lnewrs}
\end{equation}
where
\begin{equation}
%{\overset{(r) \; \; }{p'}} = {\overset{(r-s) \; \; \; }{L_{rs}}} {\overset{(s) \; \; }{p}}, \; \; \; {\rm with} \; \; \; 
{\overset{(r-s) \; \; \; }{L_{rs}}} = {\overset{(r) \; \; }{A}} {\overset{(0) \; \; \; }{L_{00}}} {\overset{(s) \; \; }{A^{-1}}},
\label{Lrsdef}
\end{equation}  
describes the Lorentz transformation from sector $s$ onto sector $r$, and where the superscript
$(r-s)$ accordingly to the $Z_3$-grading is taken modulo $3$. 

In order to provide the formulae for $Z_3$-graded boosts in explicit form we will choose the four-vectore $p_{\mu} = (p_0, \; {\bf p})$
as restricted to the plane $(0,1)$, with the three-vector ${\bf p}$ aligned along the first spatial axis.
In such a frame the Lorentz rotations reduce only to the boost in $(0,  1)$ plane, which is given by the following transformation:
\begin{equation}
\begin{pmatrix} {p'}_0 \cr {p'}_1  \end{pmatrix} =
\begin{pmatrix} {\rm ch} u & {\rm sh} u \cr {\rm sh} u & {\rm ch} u \end{pmatrix} \begin{pmatrix} p_0 \cr p_1  \end{pmatrix}, \; \; \; 
\label{oneboost}
\end{equation}
Subsequently, we get the following triplet of homogeneous transformations:   
${\overset{(0) \; \; \; }{L_{00}}}, \; {\overset{(0) \; \; \; }{L_{11}}} $ and $ {\overset{(0) \; \; \; }{L_{22}}} $:  
{\small \begin{equation}
{\overset{(0) \; \; \; }{L_{00}}} (u) = \begin{pmatrix} {\rm ch} u & {\rm sh} u \cr {\rm sh} u & {\rm ch} u \end{pmatrix}, \;   
{\overset{(0) \; \; \; }{L_{11}}} (u) = \begin{pmatrix} {\rm ch} u & j^2 {\rm sh} u \cr j {\rm sh} u & {\rm ch} u \end{pmatrix}, \;   
{\overset{(0) \; \; \; }{L_{22}}} (u) = \begin{pmatrix} {\rm ch} u & j {\rm sh} u \cr j^2 {\rm sh} u & {\rm ch} u \end{pmatrix}
\label{Lzero}
\end{equation}}
preserving respectively the bilinear forms ${\overset{ \; (r ) \; \; \; }{C_{2}}}$ (see {\ref{threeC2}).

The matrices (\ref{Lzero}) are self-adjoint:
{\small \begin{equation}
{\overset{(0) \dagger \; \; }{L_{00}}} = {\overset{(0) \; \; \; }{L_{00}}}, \; 
{\overset{(0) \dagger \; \; }{L_{11}}} = {\overset{(0) \; \; \; }{L_{11}}}, \; 
{\overset{(0) \dagger \; \; }{L_{22}}} = {\overset{(0) \; \; \; }{L_{22}}}
\label{Lzeroself}
\end{equation}}

The generalized Lorentz boosts (\ref{Lzero}) conserve the group property: the product of two Lorentz boosts acting in the $r$-th sector is a boost 
of the same type. Indeed, we see from (\ref{Lzero}) that the product of two boosts acting in the $r$-th sector ($r = 0,1,2$) looks as follows (no summation
 over $r$!): 
\begin{equation}
{\overset{(0) \; \; }{L_{rr}}} (u) \cdot {\overset{(0) \; \; }{L_{rr}}} (v) = {\overset{(0) \; \; }{L_{rr}}} (u+v).
\label{groupprop2}
\end{equation}
If we look at three fourdimensional Lorentz boost transformations on planes $(0, i), \; i=1,2,3$, the respective
set of three independent ``classical" Lorentz boosts  belonging to ${\overset{(0) \; }{L_{00}}}$
requires the introduction of three $4 \times 4$ matrices with three independent parameters $u, v, w$:
{\small \begin{equation}
\begin{pmatrix} {\rm ch} u & {\rm sh} u & 0 & 0 \cr {\rm sh} u & {\rm ch} u & 0 & 0 \cr 0 & 0 & 1 & 0 \cr 0 & 0 & 0 & 1 \end{pmatrix}, \; 
\begin{pmatrix} {\rm ch} v & 0 & {\rm sh} v & 0 \cr 0 & 1 & 0 & 0 \cr {\rm sh} v & 0 & {\rm ch} v & 0 \cr 0 & 0 & 0 & 1 \end{pmatrix}, \;  
\begin{pmatrix} {\rm ch} w & 0 & 0 & {\rm sh} w \cr 0 & 1 & 0 & 0  \cr 0 & 0 & 1 & 0 \cr {\rm sh} w & 0 & 0 & {\rm ch} w \end{pmatrix}
\label{threeboosts}
\end{equation} } 
Next, let us consider the general set of matrices (see (\ref{Lrsdef})) transforming the $s$-th sector into the $r$-th one, 
\begin{equation}
{\overset{(r) \; \; }{{p'}_{\mu}}} = ({\overset{(r-s) \; \; }{L_{rs}}})_{\mu}^{\; \; \nu} \; {\overset{(s) \; \; }{p_{\nu}}}, \; \; \; r, s = 0,1,2,
\; \; \; r \neq s.
\label{rneqsL}
\end{equation}
There are two types of such matrices: raising and lowering the $Z_3$ by $1$. For the sake of simplicity, let us firstly consider the two-dimensional
case (i.e. $\mu, \nu = 0,1$ in (\ref{rneqsL}).
The matrices $2 \times 2$ raising the $Z_3$ index $(r)$ of the generalized four-momenta 
$({\overset{(r) \; \; }{p_{\mu}}} \rightarrow  {\overset{(r+1)}{p_{\mu}}}$ are: 
{\small \begin{equation}
{\overset{(1) \; \; \; }{L_{10}}} = \begin{pmatrix} j^2 {\rm ch} u & j^2 {\rm sh} u \cr {\rm sh} u & {\rm ch} u \end{pmatrix}, \;  
{\overset{(1) \; \; \; }{L_{21}}} = \begin{pmatrix} j^2 {\rm ch} u & j {\rm sh} u \cr j {\rm sh} u & {\rm ch} u \end{pmatrix}, \;  
{\overset{(1) \; \; \; }{L_{02}}} = \begin{pmatrix} j^2 {\rm ch} u & {\rm sh} u \cr j^2 {\rm sh} u & {\rm ch} u \end{pmatrix} 
\label{Lone}
\end{equation} }
The determinants of the matrices (\ref{Lone}) are equal to $j^2$.

The matrices lowering the $Z_3$ index by one (or increasing it by $2$, what is equivalent from the point of view of the
$Z_3$-grading) are:
{\small \begin{equation}
{\overset{(2) \; \; \; }{L_{01}}} = \begin{pmatrix} j {\rm ch} u &  {\rm sh} u \cr j {\rm sh} u & {\rm ch} u \end{pmatrix}, \; 
{\overset{(2) \; \; \; }{L_{12}}} = \begin{pmatrix} j {\rm ch} u & j^2 {\rm sh} u \cr j^2 {\rm sh} u & {\rm ch} u \end{pmatrix}, \;  
{\overset{(2) \; \; \; }{L_{20}}} = \begin{pmatrix} j {\rm ch} u & j {\rm sh} u \cr {\rm sh} u & {\rm ch} u \end{pmatrix}
\label{Ltwo}
\end{equation}}
The determinants of the matrices (\ref{Ltwo}) are equal to $j$.

The above two sets of three matrices each are mutually Hermitian-adjoint:
{\small \begin{equation}
{\overset{(1) \dagger \; \; }{L_{01}}} = {\overset{(2) \; \; \; }{L_{10}}}, \;  
{\overset{(1) \dagger \; \; }{L_{12}}} = {\overset{(2) \; \; \; }{L_{21}}}, \;  
{\overset{(1) \dagger \; \; }{L_{20}}} = {\overset{(2) \; \; \; }{L_{02}}} 
\label{Lonetwo}
\end{equation} }
We recall that the superscript over each matrix ${\overset{(t) \; \; }{L_{rs}}}$ is equal to the difference of
its lower indices, i.e. $(t) = (r-s)$.

The matrices ${\overset{(1) \; \; }{L_{rs}}}$ and  ${\overset{(2) \; \; }{L_{rs}}}$ ($r,s = 0,1,2$) raising or lowering 
respectively the $Z_3$-grade of the four-momentum vectors ${\overset{(r) \; \; }{p_{\mu}}}$ do not form a Lie group.
However, together with matrices ${\overset{(0) \; \; }{L_{rs}}}$ they
can be used as building blocks in bigger $12 \times 12$ matrices forming a $Z_3$-graded generalization of the Lorentz group.  
This construction is possible due to the chain rule obeyed by these matrices, which due to the definition (\ref{Lrsdef})
display the group property. We have:
\begin{equation}
{\overset{(r-s) \; \; \; }{L_{rs}}} (p_0, p_1; \; u)  {\overset{(s-t) \; \; \; }{L_{st}}} (p_0, p_1; \; v) =  
{\overset{(r-t) \; \; \; }{L_{rt}}} (p_0, p_1; \; (u+v)).
\label{ChainLrs}
\end{equation}
In order to pass to arbitrary four-momentum vectors ${\overset{(r) \; \; }{p_{\mu}}}, \; \mu = 0,1,2,3$ one should embed the $2 \times 2$
matrices (\ref{Lone} - \ref{Ltwo}) into $4 \times 4$ matrices in a way analogous to passing from the $2 \times 2$ boost matrices 
${\overset{(0) \; \; \; }{L_{00}}}$ (see (\ref{Lzero})) to the triplet of boosts in planes $(0, i), i=1,2,3$ described by the $4 \times 4$
matrices (\ref{threeboosts}).

If we write a $Z_3$-extended four-momentum vector $( {\overset{(0) \; }{p^{\mu}}}, \;  {\overset{(1) \; }{p^{\mu}}}, \; {\overset{(2) \; }{p^{\mu}}} )^T$
as a column with $12$ entries, we can introduce three boost sectors 
${\overset{(r) }{\Lambda}}, \; (r=0,1,2)$ of the generalized $Z_3$-graded Lorentz group as $12 \times 12$ matrices as follows:
{\small \begin{equation}
{\overset{(0) }{\Lambda}} = \begin{pmatrix} {\overset{(0) \; \; \; }{L_{00}}} & 0 & 0 \cr 0 & {\overset{(0) \; \; \; }{L_{11}}} & 0 \cr 
0 & 0 & {\overset{(0) \; \; \; }{L_{22}}} \end{pmatrix} \; \; \; \; 
{\overset{(1)}{\Lambda}} = \begin{pmatrix} 0 & 0 & {\overset{(1) \; \; \; }{L_{02}}} \cr {\overset{(1) \; \; \; }{L_{10}}} & 0 & 0 \cr 
0 & {\overset{(1) \; \; \; }{L_{21}}} & 0 \end{pmatrix} \; \; \; 
{\overset{(2)}{\Lambda}} = \begin{pmatrix} 0 & {\overset{(2) \; \; \; }{L_{01}}} & 0  \cr 0 & 0 &  {\overset{(2) \; \; \; }{L_{12}}} \cr 
{\overset{(2) \; \; \; }{L_{20}}} & 0 & 0 \end{pmatrix}.
\label{Lambda123}
\end{equation}}
It should be stressed that in each of the $12 \times 12$ matrices ${\overset{(r)}{\Lambda}}, \; r=0,1,2$ the triplets of $4 \times 4$
matrices ${\overset{(r-s) \;  }{L_{rs}}}$ are obtained from the standard classical Lorentz boost by applying the definition (\ref{Lrsdef}),
i.e. each  ${\overset{(r) }{\Lambda}}$-matrix depends only on three parameters defining three independent Lorentz boosts. 
 
One can show that the matrices (\ref{Lambda123}) display the following $Z_3$-graded multiplication rules: 
\begin{equation}
{\overset{(0) }{\Lambda}}\cdot  {\overset{(0) }{\Lambda}} \subset {\overset{(0) }{\boldsymbol{\Lambda}}}, \; \; 
{\overset{(0) }{\Lambda}}\cdot  {\overset{(1) }{\Lambda}} \subset {\overset{(1) }{\boldsymbol{\Lambda}}}, \; \;  
{\overset{(0) }{\Lambda}}\cdot  {\overset{(2) }{\Lambda}} \subset {\overset{(2) }{\boldsymbol{\Lambda}}}, \; \;  
{\overset{(1) }{\Lambda}}\cdot  {\overset{(1) }{\Lambda}} \subset {\overset{(2) }{\boldsymbol{\Lambda}}}, 
\label{Z3Lambda1}
\end{equation}
\begin{equation}
{\overset{(2) }{\Lambda}}\cdot  {\overset{(2) }{\Lambda}} \subset {\overset{(1) }{\boldsymbol{\Lambda}}}, \; \; \; \; \
{\overset{(1) }{\Lambda}}\cdot  {\overset{(2) }{\Lambda}} \subset {\overset{(0) }{\boldsymbol{\Lambda}}}, \; \; \; \; \; 
{\overset{(2) }{\Lambda}}\cdot  {\overset{(1) }{\Lambda}} \subset {\overset{(0) }{\boldsymbol{\Lambda}}}.
\label{Z3Lambda2}
\end{equation}
where ${\overset{(r) }{\boldsymbol{\Lambda}}} \; \; (r=0,1,2)$ denote the$Z_3$-graded sectors of the full set of $12 \times 12$ matrix
Lorentz group which includes also the $Z_3$-graded $O(3)$ spatial rotations. 

The multiplication table (\ref{Z3Lambda1}-\ref{Z3Lambda2}) with the $Z_3$-graded structure can be described in a compact way 
using the bold-face symbols ${\overset{(r) }{\boldsymbol{\Lambda}}}$  as follows: 
\begin{equation}
{\overset{(r) }{\boldsymbol{\Lambda}}}\cdot  {\overset{(s) }{\boldsymbol{\Lambda}}} 
\subset {\overset{(r+s) \mid_3 }{\boldsymbol{\Lambda}}}, \; \; {\rm with} \; \; r, s,.. = 0,1,2, \; \; (r+s) \; {\rm taken \; modulo} \; 3.
\label{Multgrade}
\end{equation}

The construction of $Z_3$-graded $O(3)$ rotations completing the $Z_3$-graded boosts ${\overset{(r) }{\boldsymbol{\Lambda}}}$ is as follows.
Let us denote by $R_i$ the usual space rotation around the $i$-th axis, represented as a $3 \times 3$ matrix. When incorporated
into the four-vector representation of the Lorentz group, it becomes a sub-matrix of a $4 \times 4$ Lorentzian matrix according
to the formula ${\overset{(0)}{R_i}} = \begin{pmatrix} 1 & 0 \cr 0 & R_i \end{pmatrix} $. The $Z_3$-graded space rotations supplementing 
the $Z_3$-graded boosts (\ref{Lambda123}) are constructed as the following $12 \times 12$ matrices:
\begin{equation}
{\overset{(0)}{{\cal{R}}_i}} =  {\mbox{l\hspace{-0.55em}1}}_{3} \otimes  {\overset{(0)}{R_i}}, \; \; \; 
{\overset{(1)}{{\cal{R}}_i}} =  Q^{\dagger}_3 \otimes  {\overset{(0)}{R_i}}, \; \; \; 
{\overset{(2)}{{\cal{R}}_i}} =  Q_3 \otimes  {\overset{(0)}{R_i}},
\label{Z3rotations}
\end{equation}
where the choice of the colour generators $Q^{\dagger}_3$ and $Q_3$ is consistent with the colour Dirac equations (\ref{EPtogether}-{\ref{Gammafirst}).

%imposed by the $Z_3$-graded structure of the generalized $Z_3$-graded boosts (\ref{Lambda123});

The $Z_3$-graded infinitesimal
generators of the Lorentz boosts can be obtained by considering the matrices  ${\overset{(r)}{\Lambda}}$ with infinitesimal boost
parameters (i.e. taking the differential) what amounts to the replacements of the entries ${\rm sh} u $ by 1, and of all other entries, 
${\rm ch} u$ and $1$ alike, by $0$.   
The resulting $12 \times 12$ matrices are the Lie algebra generators of the generalized Lorentz boosts, which we shall denote
as $ {\overset{(r)}{K_i}}, \; r = 0,1,2$. By taking their commutators we obtain the $Z_3$-graded generators of the space rotations
$(r+s)$ modulo $3$):
\begin{equation}
[ {\overset{(r)}{K_i}}, {\overset{(s)}{K_j}} ] = - \epsilon_{ijk} {\overset{(r+s)}{J_k}}
\label{Kijcomm}
\end{equation}
In such a way we obtain the full set of generators of the $Z_3$-graded Lorentz algebra
which satisfy the following commutation relations:
\begin{equation}
\begin{split}
&\left[ J^{(r)}_i, J^{(s)}_k \right] \!\! =\!\! \epsilon_{ikl} J^{(r+s)}_l, \; \; 
  \left[ J^{(r)}_i, K^{(s)}_k \right]\!\! =\!\! \epsilon_{ikl} K^{(r+s)}_l, 
	\\
 &\left[ K^{(r)}_i, K^{(s)}_k \right]\!\! =\!\! - \epsilon_{ikl} J^{(r+s)}_l.
\end{split}
\label{modulocommAA}
\end{equation}
which were firstly introduced and studied in (\cite{RKJL2019}).

Let us consider ${\overset{(r) }{\Lambda}}$ as $3 \times 3$ matrices, with their matrix elements represented by $4 \times 4$ blocks.
${\overset{(t) }{L}}_{rs}$  (see (\ref{rneqsL}))

The matrices ${\overset{(0) }{\Lambda}}$  are Hermitian by virtue of formula (\ref{Lzeroself}), while 
$({\overset{(1) }{\Lambda}})^{\dagger} = {\overset{(2) }{\Lambda}}$ or equivalently, $({\overset{(2) }{\Lambda}})^{\dagger} = {\overset{(1) }{\Lambda}}$
as a result of formula (\ref{Lonetwo}).

Any group structure of $12 \times 12$ matrices is preserved under the similarity transformations, 

$\Lambda  \rightarrow {\tilde{\Lambda}} = {\cal{U}} \Lambda {\cal{U}}^{-1}$, but the above Hermitian properties of 
$\Lambda$-matrices are conserved only if the transformation matrices are unitary. The proof is immediate: let us denote by 
${\cal{U}}= U \otimes \; {\mbox{l\hspace{-0.55em}1}}_{4}  $ a $12 \times 12$ matrix obtained as a tensor product of a $3 \times 3$ complex valued
$U$ matrix by the unit $4 \times 4$ matrix \; ${\mbox{l\hspace{-0.55em}1}}_{4}$. 
(Obviously, ${\cal{U}}^{\dagger} = U^{\dagger} \otimes {\mbox{l\hspace{-0.55em}1}}_{4}$).
\noindent
Let us firstly define ${\overset{(0) }{\Lambda}} \rightarrow {\cal{U}} {\overset{(0) }{\Lambda}} {\cal{U}}^{-1}$
and impose the Hermiticity conditions on the transformed matrices ${\cal{U}} {\overset{(0) }{\Lambda}} {\cal{U}}^{-1} $. Because the matrix 
$ {\overset{(0) }{\Lambda}}$ is Hermitean, we have
\begin{equation}
 \left( {\cal{U}} {\overset{(0) }{\Lambda}} {\cal{U}}^{-1} \right)^{\dagger}  
= ({\cal{U}}^{-1})^{\dagger} {\overset{(0) }{\Lambda}}^{\dagger} {\cal{U}}^{\dagger} = {\cal{U}} {\overset{(0) }{\Lambda}} {\cal{U}}^{-1},
\label{UnitaryU}
\end{equation}
the matrix ${\cal{U}} {\overset{(0) }{\Lambda}} {\cal{U}}^{-1}$ is Hermitean, too, if the similarity matrices are {\it unitary}, 
if i.e. ${\cal{U}}^{\dagger} = {\cal{U}}^{-1}$,  according to the formula ${\cal{U}} = U \otimes {\mbox{l\hspace{-0.55em}1}}_{4}$ 
it follows that $U^{\dagger} = U^{-1}$. Hermitian conjugation relations
between the matrices ${\overset{(1) }{\Lambda}}$ and ${\overset{(2) }{\Lambda}}$ are also preserved after similarity transformation 
if the similarity matrices obey the same unitarity condition ${\cal{U}}^{\dagger} = {\cal{U}}^{-1}$.

In this way we introduced the symmetry $SU(3)$ acting on the vector representation of the $Z_3$-graded Lorentz group. The $3 \times 3$ matrices $U$ 
appearing in the $12 \times 12$ matrices ${\cal{U}}$ during the unitary similarity transformations leave the $4 \times 4$ Lorentzian blocks unaffected, 
in agreement with the well known ``no-go theorems'' by Coleman and Mandula and O'Raifeartaigh (\cite{Colemandula}, \cite{Raifeartaigh}).

We point out that in order to obtain the entire $Z_3$-graded  Lorentz group
we should add as well the $Z_3$-graded extension of space rotations, also represented as $12 \times 12$ matrices, with building blocks .
 made of $4 \times 4$ matrices, just like the $Z_3$-graded boosts. 
% It is easy to see that due to the fact that spatial coordinates in 
%in three replicas of momentum $4$-space are equal and identical with the ordinary $3$-dimensional space, the space rotations should
%be the same in all sectors. However, 
As in the case of Lorentz boosts, besides the rotations that leave the transformed $3$-momentum in the same sector, 
one gets also$12 \times 12$ matrices with non diagonal $4 \times 4$ entries, which map one of the $Z_3$-graded sectors onto another one. 

We conclude that the full set of $Z_3$-graded $O(3)$ subgroup elements can be represented by $12 \times 12$ matrices and incorporated
in the $Z_3$-graded Lorentz group.

In this Section we were considering the vectorial realizations of the $Z_3$-graded Lorentz group which can be also extended 
to the realizations of $Z_3$-graded Poincar\'e algebra (see also (\cite{Kerner2019B}))
In the next two sections we will present our main result: how the sextet of the colour Dirac matrices $\Gamma^{\mu}$ appears in the construction 
of faithful spinorial $72 \times 72$ matrix representation of the $Z_3$-graded Lorentz algebra. In such a way we will be able to incorporate all internal 
symmetries of quark sector appearing in the Standard Model into a group-theoretical framework. 

\section{$Z_3$-graded generalized Lorentz algebra and its spinorial matrix realization}
\vskip 0.3cm
The $12 \times 12$ matrices $\Gamma^{\mu}$ (see (\ref{Gammasbig})) appearing in the coloured Dirac equation (\ref{Gammasecond}), 
(see also \ref{Gammasbig}) are linked with the $Z_3$-graded generalization of classical Lorentz symmetries.
 In particular, the $Z_3 \otimes Z_2 \otimes Z_2$ structure of $\Gamma^{\mu}$-matrices implies that due to the identities 
$(Q_a)^3 = (Q_a^{\dagger})^3 = {\mbox{l\hspace{-0.55em}1}}_{3}, \; \; B^3 = (B^{\dagger})^3 = {\mbox{l\hspace{-0.55em}1}}_{3}$ 
and $(\sigma_i)^2 = {\mbox{l\hspace{-0.55em}1}}_{2}$
their sixth powers are proportional to the unit matrix $ \; {\mbox{l\hspace{-0.55em}1}}_{12}$ (see also (\ref{dettwelve})). 

Let us first derive the $Z_3$-graded Lorentz algebra, which follows from the covariance properties of the colour Dirac equation (\ref{Gammasecond}).

The two standard commutators of $\Gamma^{\mu}$ matrices, namely 
\begin{equation}
 J_i = \frac{i}{2} \; \epsilon_{ijk} \left[ \Gamma^j, \Gamma^k \right], \; \; \; K_l = \frac{1}{2} \; \left[ \Gamma_l, \Gamma_0 \right] \; 
\label{ThreeJK}
\end{equation}
provide only the first step towards the construction of the generators of a $Z_3$-graded Lorentz algebra.
Surprisingly, one can check that the generators $\left( J^{(0)}_i, \; K^{(0)}_l \right)$ 
satisfying the standard $D=4$ L                                                                                                                                                                                                                                                                                                                        orentz algebra relations 
\begin{equation}
[ \; {\overset{(0) }{K}}_i, \; {\overset{(0) }{K}}_k \; ]  = - \epsilon_{ikl} {\overset{(0) }{J}}_l, \; \; \; \; \; 
[ \; {\overset{(0) }{J}}_i, \; {\overset{(0) }{K}}_k \; ]  = \epsilon_{ikl} {\overset{(0) }{K}}_l, \; \; \; \; \;%
\label{modulocomm0}
\end{equation}
\begin{equation}
[ \; {\overset{(0) }{J}}_i, \; {\overset{(0) }{J}}_k \; ] = \epsilon_{ikl} {\overset{(0) }{J}}_l.
\label{modulocomm1}
\end{equation}
can be defined  by {\it double} commutators of $12 \times 12$ matrices $J_i, K_l$ as follows: 
\begin{equation}
\begin{split}
&\left[ J_i, \left[J_j, J_k \right] \right] = \left( \delta_{ij}\delta_{kl} - \delta_{ik} \delta_{jl} \right) \; J^{(0)}_l,
\\
&\left[ K_i, \left[K_j, K_k \right] \right] = \left( \delta_{ij}\delta_{kl} - \delta_{ik} \delta_{jl} \right) \; K^{(0)}_l \; .
\end{split}
\label{triplecomm}
\end{equation}
Using the definition of standard colour $\Gamma^{\mu}$-matrices (\ref{Gammasbig}) 
and substituting it in (\ref{ThreeJK} and \ref{triplecomm}), we get  
\begin{equation}
\begin{split}
&J_i = - \frac{i}{2} \; Q_2^{\dagger} \otimes {\mbox{l\hspace{-0.55em}1}}_{2} \otimes \sigma_i, \; \; \; K_l = 
- \frac{1}{2} \; Q_1 \otimes \sigma_1 \otimes \sigma_l,
\\
&J^{(0)}_i = - \frac{i}{2} \; {\mbox{l\hspace{-0.55em}1}}_{3} \otimes \; {\mbox{l\hspace{-0.55em}1}}_{2} \otimes \sigma_i, \; \; \; 
K^{(0)}_l = - \frac{1}{2} \; {\mbox{l\hspace{-0.55em}1}}_{3}  \otimes \sigma_1 \otimes \sigma_l.
\label{NewJK}
\end{split}  
\end{equation}
 In order to introduce the $Z_3$-graded Lorentz algebra
\begin{equation}
{\cal{L}} = L^{(0)} \oplus L^{(1)} \oplus L^{(2)}
\label{Lcal}
\end{equation}
where $L^{(0)}\!\! =\!\! ( J^{(0)}_i, K^{(0)}_j ),$ \; \;   $L^{(1)}\!\! =\!\! ( J^{(1)}_i, K^{(1)}_j ),$ \; \; 
$ L^{(2)}\!\! =\!\! ( J^{(2)}_i, K^{(2)}_j ),$ 
one should supplement the relations (\ref{triplecomm}) by the pairs of other possible double commutators: 
\begin{equation}
\begin{split}
&\left[ J_i, \left[J_j, K_k \right] \right] = 
\left( \delta_{ij}\delta_{kl} - \delta_{ik} \delta_{jl} \right) \; K^{(2)}_l,
\\
&\left[ K_i, \left[K_j, J_k \right] \right] = \left( \delta_{ij}\delta_{kl} - \delta_{ik} \delta_{jl} \right) \; J^{(1)}_l,
\end{split}
\label{tripcommextra}
\end{equation}
In particular,  besides the representation (\ref{NewJK}) we get the following realizations:
\begin{equation}
J^{(1)}_l = -\frac{i}{2} \; Q_3 \otimes \; {\mbox{l\hspace{-0.55em}1}}_{2} \otimes \sigma_l, \; \; \; \; K^{(2)}_i 
= - \frac{1}{2} \; Q_3^{\dagger} \otimes \sigma_1 \otimes \sigma_i.
\label{J1K2}
\end{equation}
The three-linear double commutators in (\ref{triplecomm}) and (\ref{tripcommextra}) are related with $Z_3$-grading;
when taken into account, the full set of $Z_3$-graded relations defining the $Z_3$-graded Lorentz algebra introduced in \cite{RKJL2019}
(where $ r, s, = 0,1,2, \; r+s$ are taken modulo $3$), results in the following set of commutation relations (\cite{RKJL2019}):  
\begin{equation}
\begin{split}
&\left[ J^{(r)}_i, J^{(s)}_k \right] \!\! =\!\! \epsilon_{ikl} J^{(r+s)}_l, \; \; 
  \left[ J^{(r)}_i, K^{(s)}_k \right]\!\! =\!\! \epsilon_{ikl} K^{(r+s)}_l, 
	\\
 &\left[ K^{(r)}_i, K^{(s)}_k \right]\!\! =\!\! - \epsilon_{ikl} J^{(r+s)}_l.
\end{split}
\label{modulocommB}
\end{equation}

From the commutators 
$[ K^{(1)}, K^{(1)} ] \simeq J^{(2)}$ and $[ J^{(1)}, J^{(1)} ] \simeq J^{(2)}$ 
one gets the realization of remaining generators of ${\cal{L}}$,
\begin{equation}
J^{(2)}_i = - \frac{i}{2} \; Q_3^{\dagger} \otimes {\mbox{l\hspace{-0.55em}1}}_{2} \otimes \sigma_i, \; \; \; 
K_m^{(1)} = - \frac{1}{2} \; Q_3 \otimes \sigma_1 \otimes \sigma_m.
\label{J2K1}
\end{equation}

The formulae (\ref{NewJK}), (\ref{J1K2}) and (\ref{J2K1}) describe the spinorial realization of the Lie algebra ${\cal{L}}$
which is implied by the choice (\ref{Gammasbig}) of matrices $\Gamma^{\mu}$. Let us introduce a unified notation englobing 
all possible choices of $\Gamma^{\mu}$-matrices ($A \neq B$)
\begin{equation}
\Gamma^0_{(A; \alpha)} = I_A \otimes \sigma_{\alpha} \otimes \sigma^0, \; \; \; \Gamma^i_{(B; \beta)} 
= I_B \otimes (i \sigma_{\beta}) \otimes \sigma^i,
\label{GammawithI}
\end{equation}
where $I_0 = {\mbox{l\hspace{-0.55em}1}}_{3}$, $I_A $ with $A=1,2,...,8$ are given in (\ref{IAdef}), and  $\alpha, \beta = 2,3$ 
but $\{ \sigma_{\alpha}, \; \sigma_{\beta} \}_{+} = 0 $ i.e. we always have either $\alpha = 2, \; \beta=3$
or $\alpha = 3, \beta=2$.  

The choice $\alpha =1$
is not present in the formula (\ref{GammawithI})  because it is reserved for the description of symmetry generators ${\cal{L}}$ 
(see (\ref{J1K2}), (\ref{J2K1})). Further, eight colour $3 \times 3$ matrices $I_A$ ( $A=1,2,...8$ ) with the multiplication rules given 
in Table 1 span the ternary basis of the $SU(3)$ algebra (\cite{Kac1994}, Sect. $8$).  

The characteristic feature of ``colour'' $\Gamma$-matrices is that the $3 \times 3$ matrices $I_A$ appearing as the first
tensorial factors in (\ref{GammawithI}) are {\it different} for temporal and spatial components of the matrix-valued $4$-vector $\Gamma^{\mu}$.
We see that the choice of the colour factor in (\ref{GammawithI}) depends on two sets of values of the four-vector 
index: $\mu = 0$ or $\mu= i$ where $i=1,2,3$. This property can be interpreted as the entanglement of colour and Lorentz symmetry degrees of freedom.
\noindent
In the notation (\ref{Gammaexplicit}) basic $\Gamma$-matrices (\ref{Gammasbig}) derived in Section $3$ can be denoted as 
\begin{equation}
\Gamma^0_{(8, 3)} = B^{\dagger} \otimes \sigma_3 \otimes {\mbox{l\hspace{-0.55em}1}}_{2}, \; \; \; \; 
\Gamma^i_{(2; 2)} = Q_2 \otimes (i \sigma_2) \otimes \sigma^i. 
\label{Gammaexplicit}
\end{equation}
In order to get a closed formula for the adjoint action 
${\cal{S}}^{(0)} \Gamma^{\mu} [{\cal{S}}^{(0)}]^{-1}$  of classical spinorial Lorentz group (see (\ref{modulocomm0}, \ref{modulocomm1}), 
where $a^i, b^k, \; \; (i, k = 1,2,3)$ are the six real $SL(2, {\bf C})$ Lie group parameters
\begin{equation}
S^{(0)} = \exp \; \left( a^i K^{(0)}_i +  b^k J^{(0)}_k\right)
 \label{Sexp}
\end{equation}
we should introduce the following pairs of 
 $\Gamma^{\mu}$-matrices 
\begin{equation}
\Gamma^{\mu} = ( \Gamma^{i}_{(A ; 2)}, \; \Gamma^{0}_{(B;3)}) \; {\rm  and} \;  {\tilde{\Gamma}}^{\mu} = 
(\Gamma^{i}_{(B; 2)}, \; \Gamma^{0}_{(A; 3)}),
\label{Gamwithtilde}
\end{equation}
 where we have chosen in (\ref{Gammaexplicit}) $\alpha = 3$ and $\beta = 2$.
Although for {\it any} choice of the first factor $I_A$ in  $\Gamma_{(A; \alpha)}^{\mu}$'s 
(see  \ref{GammawithI}) 
we have 
\begin{equation}
\left[ J_i^{(0)}, \Gamma^{j}_{(A; \alpha)} \right] = 
\epsilon_{ijk} \Gamma^{k}_{(A; \alpha)}, \; \; \;  
\left[ J_i^{(0)}, \Gamma^{0}_{(A; \alpha)} \right] =  0,
\label{JGamma0}
\end{equation}
the boosts $K_i^{(0)}$ (see (\ref{NewJK})) act covariantly only on doublets $\left( \Gamma^{\mu}, {\tilde{\Gamma}}^{\mu} \right)$,
with $(A \neq B)$, because only for such a choice we can get the closure of commutation relations:
$$[ K_i^{(0)}, \Gamma^{j}_{(A; 2)} ] = \delta^j_i \;  \Gamma^{0}_{(A; 3)}, \;  \; \;  
[ K_i^{(0)}, \Gamma^{0}_{(B; 3)} ] = \Gamma^{i}_{(B; 2)},$$
\begin{equation}
[ K_i^{(0)}, \Gamma^{j}_{(B; 2)} ] = \delta_i^j \; \Gamma^{0}_{(B; 3)}, \; 
[K_i^{(0)}, \Gamma^{0}_{(A; 3)} ] = \Gamma^{i}_{(A; 2)}.
\label{Kongamma2}
\end{equation}
It follows from (\ref{JGamma0}), (\ref{Kongamma2}) that the standard Lorentz covariance requires the pair of coloured Dirac equations 
described by the {\it doublet} 
$(\Gamma^{\mu}, \; {\tilde{\Gamma}}^{\mu})$ of coloured Dirac matrices (see \ref{Gamwithtilde}),  which we shall call ``{\it Lorentz doublets}''.  
In particular, the $\Gamma^{\mu}$ matrices (\ref{Gammasbig}) from Sect. 3 should be supplemented by the following Lorentz doublet partner:
\begin{equation}
{\tilde{\Gamma}}^0 = \Gamma^{0}_{(2; 3)} = Q_2 \otimes \sigma_3 \otimes {\mbox{l\hspace{-0.55em}1}}_{2},
\; \; \; {\tilde{\Gamma}}^i = \Gamma^{i}_{(8; 2)} = B^{\dagger} \otimes (i \sigma_2) \otimes \sigma^{i}.
\label{Pairgammas}
\end{equation}
Further we will show that the Lorentz doublets of $\Gamma^{\mu}$-matrices required by the standard Lorentz covariance 
can be useful for the description of weak isospin (flavour) doublets
of the $SU(2) \times U(1)$ electroweak symmetry. In such a way one can show that the internal symmetries
$SU(3) \times SU(2) \times U(1)$ of Standard Model are linked with the presence of standard Lorentz covariance which
generates three $24$-component Lorentz doublets of colour Dirac spinors.

Next, we will show that in order to obtain the closure of the faithful action of generators 
$(J_k^{(s)}, \; K_m^{(s)})$ ($s = 0,1,2$) which describe the $Z_3$-graded spinorial transformations of matrices $\Gamma^{\mu}$ ,
we need two Hermitean-conjugate sextets ($\Gamma^{\mu}_{(a)},  \Gamma^{\mu}_{(\dot{a})} = (\Gamma^{\mu}_{(a)})^{\dagger})  \; (a = 1,2,...6)$ 
of coloured $12 \times 12$ Dirac matrices. We will also show that the sextet $\Gamma^{\mu}_{(a)}$ defines three Lorentz doublets
needed for the implementation of classical Lorentz covariance.
\section{Irreducible spinorial representation of $Z_3$-graded Lorentz algebra and
colour $\Gamma^{\mu}$ matrices as its module}
\vskip 0.1cm
\subsection{Sextet of $\Gamma^{\mu}$-matrices following from the $Z_3$-graded Lorentz covariance}
\vskip 0.2cm
Let us choose $( J_k^{(1)}, K_m^{(1)} ) $  as given by Eqs. (\ref{J1K2}), (\ref{J2K1}), and assume that 
 $\Gamma^{\mu}_{(1)}$ describes the $\Gamma^{\mu}$-matrix (\ref{Gammasbig}) and respectively, ${\tilde{\Gamma}}^{\mu}$, 
its doublet partner (\ref{Pairgammas}).  
By calculating the multicommutators of $\left( J^{(1)}_i, K^{(1)}_l \right) \in L^{(1)}$  with 
the set $\Gamma^{\mu}_{(a)}, \; (a=1,2...6)$,  we will show that the following sextet of $\Gamma$-matrices which break the Lorentz covariance
is closed under the action of $L^{(1)}$ :
\begin{equation}
\begin{split}
&\Gamma^{\mu}_{(1)} = \left( \Gamma^0_{(8;3)}, \;  \Gamma^i_{(2;2)} \right); \; \; \; \; 
\Gamma^{\mu}_{(4)} = \left( \Gamma^0_{(8;2)}, \; \Gamma^i_{(2;3)} \right); 
\\
&\Gamma^{\mu}_{(2)} =  \left( \Gamma^0_{(2;2)},  \; \Gamma^i_{(4;3)} \right); \; \;  \; \; 
\Gamma^{\mu}_{(5)} = \left( \Gamma^0_{(2;3)},  \; \Gamma^i_{(4;2)} \right);
\\
&\Gamma^{\mu}_{(3)} =  \left( \Gamma^0_{(4;3)},  \; \Gamma^i_{(8;2)} \right); \; \; \; \; 
\Gamma^{\mu}_{(6)} = \left( \Gamma^0_{(4;2)},  \;  \Gamma^i_{(8;3)} \right).
\end{split}
\label{sixGammas}
\end{equation}
It is easy to see that from the six components of the sextet (\ref{sixGammas}) one can construct as well the set of six $\Gamma^{\mu}$-matrices
$\Gamma^{\mu}_{(A; \alpha)}, \; A=2,4,8$ and $\alpha= 2,3$, which can be described as well as three Lorentz doublets (\ref{Gamwithtilde}),
with $(A,B) = (2,8), (2,4)$ and $(4,8)$.. More explicitly, 
\begin{equation}
\left( \Gamma^0_{(A;\alpha)} = I_A \otimes \sigma_{\alpha} \otimes \; {\mbox{l\hspace{-0.55em}1}}_{2}, \; 
\Gamma^i_{(B;\beta)} = I_B \otimes (i \sigma_{\beta}) \otimes \sigma^i \right), 
\label{GammaIexp}
\end{equation} 
where $ (I_A, I_B = Q_2, \; Q_1^{\dagger}, \; B^{\dagger})$ and $\alpha = 2,3.$
The construction of the $Z_3$-graded Lorentz generators $J_k^{(r)}, \; K_m^{(s)}$ employed as first tensorial factor the $3 \times 3$
matrices \;  ${\mbox{l\hspace{-0.55em}1}}_{3}$ \; for $r, s =0$, $Q_3$ for $r, s =1$ and $Q_3^{\dagger}$ for $r, s = 2$. From the remaining six
generators of the $SU(3)$ Lie algebra in the Kac basis, only three do appear in the sextet (\ref{GammaIexp}). In order to implement the
 full $SU(3)$ colour symmetry, the remaining matrices $Q_1, Q_2^{\dagger}$ and $B$ should be included in the module on which acts via commutation 
the spinorial representation of the $Z_3$-graded Lorentz algebra. This means that the following sextet should be also
taken into consideration, obtained by replacing the matrices $I_A$ by their complex conjugates, and keeping the remaining tensorial
factors unchanged:
\begin{equation}
 \Gamma^0_{({\bar{A}};\alpha)} = {\bar{I}}_A \otimes \sigma_{\alpha} \otimes \sigma^{0}, \; \; {\rm where} \; \; 
{\bar{I}}_A = ( Q_1, \; Q_2^{\dagger}, \; B)
\label{GammmaIexp}
\end{equation}
 The realization of $L^{(2)}$ sector acting on colour $\Gamma^{\mu}$ matrices is obtained by introducing the Hermitean-conjugate sextet 
 $\Gamma^\mu_{(\dot{a})} = (\Gamma^\mu_{(a)})^{\dagger}$
\begin{equation}
\begin{split}
&\Gamma^{\mu}_{({\dot{1}})} = (\Gamma^{\mu}_{(1)})^{\dagger} = \left( \Gamma^0_{(7;3)}, \; \Gamma^i_{(5;2)} \right); \; \; \; \;  
\Gamma^{\mu}_{({\dot{4}})} = (\Gamma^{\mu}_{(4)})^{\dagger} = \left( \Gamma^0_{(7;2)}, \;  \Gamma^i_{(5;3)} \right); 
\\
&\Gamma^{\mu}_{({\dot{2}})} =  (\Gamma^{\mu}_{(2)})^{\dagger} = \left( \Gamma^0_{(5;2)},  \; \Gamma^i_{(1;3)} \right); \; \; \; \;   
\Gamma^{\mu}_{({\dot{5}})} = (\Gamma^{\mu}_{(5)})^{\dagger} = \left( \Gamma^0_{(5;3)},  \; \Gamma^i_{(1;2)} \right);
\\
&\Gamma^{\mu}_{({\dot{3}})} =  (\Gamma^{\mu}_{(3)})^{\dagger} = \left( \Gamma^0_{(1;3)}, \; \Gamma^i_{(7;2)} \right); \; \; \; \; 
\Gamma^{\mu}_{({\dot{6}})} = (\Gamma^{\mu}_{(6)})^{\dagger} = \left( \Gamma^0_{(1;2)},  \; \Gamma^i_{(7;3)} \right),
\end{split}
\label{sixGammasdag}
\end{equation}
linearly related with the tilded $\Gamma$-matrices  
${\tilde{\Gamma}}^\mu_{(\dot{a})} = ({\tilde{\Gamma}}^\mu_{(a)})^{\dagger}$ which are 
required by standard Lorentz covariance described by the grade $0$ sector  $L^{(0)}$ 
(see (\ref{modulocomm0}), (\ref{modulocomm1}) and (\ref{Lcal}), (\ref{tripcommextra})).

\subsection{Lorentz doublets and classical Lorentz symmetry - sector $L^{(0)}$}
\vskip 0.2cm
The action of zero-grade rotation generators $J^{(0)}_i$ on coloured matrices $\Gamma^{\mu}$ is described by the eq. (\ref{JGamma0}).
In particular, the space rotations leave the temporal component
$\Gamma^0_{(A; \alpha)}$ invariant, and transform the space components as the coordinates of a $D=3$ three-vector,
while the commutators of boosts $K_i^{(0)}$ with  $( \Gamma^0_{(A, \alpha)}, \; \Gamma^i_{(B; \beta)} $ generate new $\Gamma^{\mu}$-matrices
%with both time ($\mu=0$) as well as space $(\mu = 1,2,3)$ indices, 
which permit to introduce the ``Lorentz partners''.

Let us start with the first ``standard'' choice of colour $\Gamma^{\mu}$-matrices (see \ref{Gammaexplicit})
\begin{equation}
\Gamma^{\mu}_{(1)} = \left( \Gamma^0_{(8;3)}, \; \Gamma^i_{(2;2)} \right) = 
\left( B^{\dagger} \otimes \sigma_3 \otimes {\mbox{l\hspace{-0.55em}1}}_2,
\; Q_2 \otimes (i \sigma_2) \otimes \sigma^i \right).
\label{Gammexpnum}
\end{equation}

When iterated, the commutators of boosts $K^{(0)}_i$ with $\Gamma^{\mu}_{(1)}$-matrices yields the following result:
\begin{equation}
[ K_i^{(0)} , \Gamma^0_{(8 ; 3)} ]  = \Gamma^i_{(8;2)} = \Gamma^i_{(4)}, \; \; \; 
[ K_i^{(0)} , \Gamma^j_{(2; 2)} ]  = \delta^j_i \; \Gamma^0_{(2; 3)} = \delta^j_i \; \Gamma^0_{(4)},
\label{Kongamma3}
\end{equation} 
\begin{equation}
[ K_i^{(0)} , \Gamma^0_{(2; 3)} ]  = \Gamma^i_{(2;2)} = \Gamma^i_{(1)}, \; \; \; 
[ K_i^{(0)} , \Gamma^j_{(8; 2)} ]  = \delta^j_i \; \Gamma^0_{(8; 3)} = \delta^j_i \; \Gamma^0_{(1)}.
\label{Kongamma4}
\end{equation} 
Apparently, we obtain a classical Lorentz doublet $\left( \Gamma^{\mu}_{(1)}, \; {\tilde{\Gamma}}^{\mu}_{(1)} \right)$, where
${\tilde{\Gamma}}^{\mu}_{(1)} = \left( \Gamma^0_{(5)}, \; \Gamma^i_{(3)} \right)$, (see \ref{sixGammas}). It appears that in an analogous manner 
one can introduce classical Lorentz doublets for each colour $\Gamma^{\mu}$-matrix listed in (\ref{sixGammas}) by adding to $\Gamma^{\mu}_{(a)}$
the multiplet ${\tilde{\Gamma}}^{\mu}_{(a)} = \left( \Gamma^0_{(b)}, \; \Gamma^i_{(c)} \right) $, where $b = a+4 \; (mod \; 6)$ and $c = a+2 \; (mod \; 6)$.

After the calculation of commutators of the boosts $K_i^{(0)}$ with all $\Gamma^{\mu}$-matrices which appear in (\ref{sixGammas}),
the following sextet of Lorentz doublets $(\Gamma^{\mu}_{(a)}, {\tilde{\Gamma}}^{\mu}_{(a)}), \; (a = 1,2,...,6)$ is obtained:

$$\Gamma^{\mu}_{(1)} = ( \Gamma^{0}_{(8;3)}, \Gamma^i_{(2,2)} ) = ( {\tilde{\Gamma}}^{0}_{(3)}, {\tilde{\Gamma}}^i_{(5)} ), \; \; \; \; \; \; \; \; 
{\tilde{\Gamma}}^{\mu}_{(1)} = ( \Gamma^0_{(2;3)}, \; \Gamma^i_{(8;2)} ) = ( \Gamma^0_{(5)}, \; \Gamma^i_{(3)} ); $$
$$\Gamma^{\mu}_{(2)} = ( \Gamma^{0}_{(2;2)}, \Gamma^i_{(4,3)} ) = ( {\tilde{\Gamma}}^{0}_{(4)}, {\tilde{\Gamma}}^i_{(6)} ), \; \; \; \; \; \; \; \; 
{\tilde{\Gamma}}^{\mu}_{(2)} = ( \Gamma^0_{(4;2)}, \; \Gamma^i_{(2;3)} ) = ( \Gamma^0_{(6)}, \; \Gamma^i_{(4)} ); $$
\begin{equation}
\Gamma^{\mu}_{(3)} = ( \Gamma^{0}_{(4;3)}, \Gamma^i_{(8,2)} ) = ( {\tilde{\Gamma}}^{0}_{(5)}, {\tilde{\Gamma}}^i_{(1)} ), \; \; \; \; \; \; \; \; \; 
{\tilde{\Gamma}}^{\mu}_{(3)} = ( \Gamma^0_{(8;3)}, \; \Gamma^i_{(4,2)} ) = ( \Gamma^0_{(1)}, \; \Gamma^i_{(5)} ); 
\label{SextetG}
\end{equation}
$$\Gamma^{\mu}_{(4)} = ( \Gamma^{0}_{(8;2)}, \Gamma^i_{(2,3)} ) = ( {\tilde{\Gamma}}^{0}_{(6)}, {\tilde{\Gamma}}^i_{(2)} ), \; \; \; \; \; \; \; \; 
{\tilde{\Gamma}}^{\mu}_{(4)} = ( \Gamma^0_{(2;2)}, \; \Gamma^i_{(8,3)} ) = ( \Gamma^0_{(2)}, \; \Gamma^i_{(6)} ); $$
$$\Gamma^{\mu}_{(5)} = ( \Gamma^{0}_{(2;3)}, \Gamma^i_{(4,2)} ) = ( {\tilde{\Gamma}}^{0}_{(1)}, {\tilde{\Gamma}}^i_{(3)}), \; \; \; \; \; \; \; \; 
{\tilde{\Gamma}}^{\mu}_{(5)} = ( \Gamma^0_{(4;3)}, \; \Gamma^i_{(2,2)} ) =( \Gamma^0_{(3)}, \; \Gamma^i_{(1)}); $$
$$\Gamma^{\mu}_{(6)} = ( \Gamma^{0}_{(4;2)}, \Gamma^i_{(8,3)} ) = ( {\tilde{\Gamma}}^{0}_{(2)}, {\tilde{\Gamma}}^i_{(4)}), \; \; \; \; \; \; \; \; 
{\tilde{\Gamma}}^{\mu}_{(6)} = ( \Gamma^0_{(8;2)}, \; \Gamma^i_{(4,3)} ) = ( \Gamma^0_{(4)}, \; \Gamma^i_{(2)} ), $$
where we added, for the sake of completeness, the inverse formulae ${\tilde{\Gamma}}^{\mu}_{(a)} \rightarrow \Gamma^{\mu}_{(a)}$. 
%The action of ${\cal{L}}^{(0)}$ closes on the Lorentz doublets $\left[ \Gamma^{\mu}_{(a)}, \Gamma^{\mu}_{(a+4)} \right]$.
\subsection{Sextet of colour $\Gamma^{\mu}$-matrices and representations of $Z_3$-graded Lorentz algebra - 
sector $L^{(1)}= J^{(1)}_i \oplus K^{(1)}_j$}
\vskip 0.2cm
Calculating the commutators of matrices $\Gamma^{\mu}_{(a)}$ with the generators ($J_i^{(0)}, \; K_m^{(0)} \in L^{(0)}$   
was rather easy, because the only non-commuting tensorial factors were the
$3 \times 3$ ``colour'' matrices, while in the remaining two $Z_2 \times Z_2$ factors matrices $\sigma_i$ commuted with the
$2 \times 2$ unit matrices. 
However, when we consider the commutators of the operators  $ ({\overset{(r)}{J}}_i, {\overset{(s)}{K}}_m), \; \; r, s = 1,2$ 
with two colour Dirac matrices $\Gamma^{\mu}_{(1)} , \Gamma^{\mu}_{(2)}$ defined above, we generate subsequently new commutators 
we need to calculate.  

Let us observe how the new set (\ref{sixGammas})  of $\Gamma^{\mu}$-matrices is produced. 
Calculating the commutators with the grade $1$ generators we use the multiplication rule for tensor products of matrices:
\begin{equation}
(a \otimes b) \cdot (c \otimes d) = (a \cdot c) \otimes (b \cdot d), 
\label{tensormult}
\end{equation}
with $a \cdot c$ and $b \cdot d$ denoting ordinary matrix multiplication. The following formula will be helpful in our calculations:
\begin{equation}
\left[ a \otimes b, c \otimes d \right] = a \cdot c \otimes \{ b, d \} - [ a, c ] \otimes d \cdot b, \; \; {\rm where} \; \; 
\{ b, d \} = b \cdot d + d \cdot b, \; \; [a,c] = a \cdot b - b \cdot a.
\label{commfact}
\end{equation}
We recall also the well known identities involving Pauli's $\sigma$-matrices:
\begin{equation} \sigma^i \sigma^j = \delta^{ij} \; {\mbox{l\hspace{-0.55em}1}}_{2} + i \; \epsilon^{ijk} \sigma^k, \; \; \; 
\{ \sigma^i , \sigma^k \} = 2  \delta^{ik} \; {\mbox{l\hspace{-0.55em}1}}_{2}
\label{sigmasmult}
\end{equation}
\subsubsection{ Grade $1$ space rotations: sector $J^{(1)}_i$ }
\vskip 0.2cm
\indent
Let us start with grade $1$ rotations acting on $\Gamma^{\mu}_{(1)} = ( \Gamma^0_{(8;3)}, \Gamma^i_{(2;2)} ) $, 
forming the $12 \times 12$ matrix valued four-vector (\ref{Gammasbig}) appearing in the colour Dirac equation (\ref{Gammasecond}):
With the use of the rules of matrix multiplication of tensor products, we arrive at the following sequences of commutators:

$$\left[ J_i^{(1)}, \; \Gamma^0_{(8; 3)} \right] = - \frac{\beta}{2} \Gamma^i_{(2; 3)} =- \frac{\beta}{2} \Gamma^i_{(4)};$$
$$\left[ J_i^{(1)}, \; \Gamma^k_{(2; 3)} \right] = - \frac{1}{2} \epsilon_{i \; \; \; m}^{\; \; k} \Gamma^m_{(4;3)} + 
\frac{\alpha}{2} \delta_i^k \Gamma^0_{(4;3)} = - \frac{1}{2} \epsilon_{i \; \; \; m}^{\; \; k} \Gamma^m_{(2)} +
\frac{\alpha}{2} \delta_i^k \Gamma^0_{(3)};$$
$$ \left[ J_i^{(1)}, \; \Gamma^0_{(4; 3)} \right] = - \frac{\gamma}{2} \Gamma^i_{(8;3)} = - \frac{\gamma}{2} \Gamma^i_{(6)}; $$
\begin{equation}
\left[ J_i^{(1)}, \; \Gamma^k_{(4; 3)} \right] = -\frac{j^2}{2} \epsilon_{i \; \; \; m}^{\; \; k} \Gamma^m_{(8; 3)} + 
\frac{\gamma}{2} \delta_i^k \; \Gamma^0_{(8;3)} = -\frac{j^2}{2} \epsilon_{i \; \; \; m}^{\; \; k} \Gamma^m_{(6)} + 
\frac{\gamma  }{2} \delta_i^k \; \Gamma^0_{(1)};
\label{sixcomms}
\end{equation}
$$\left[ J_i^{(1)}, \; \Gamma^k_{(8; 3)} \right] = -\frac{j}{2} \epsilon_{i \; \; \; m}^{\; \; k} \Gamma^m_{(2;3)} +
\frac{\beta}{2} \delta_i^k \; \Gamma^0_{(2;3)}
-\frac{j}{2} \epsilon_{i \; \; \; m}^{\; \; k} \Gamma^m_{(4)} +\frac{\beta}{2} \delta_i^k \; \Gamma^0_{(5)};$$
$$\left[ J_i^{(1)}, \; \Gamma^0_{(2; 3)} \right] = -\frac{\alpha}{2}  \Gamma^i_{(4; 3)} = - \frac{\alpha}{2}  \Gamma^i_{(2)},$$
where we use the following shortened notation for the coefficients appearing on the right-hand side:
\begin{equation}
\alpha = j - j^2, \; \; \; \beta = j^2 -1, \; \; \gamma = 1-j, \; \; \; \; \; (j = e^{\frac{2 \pi i}{3}}). 
\label{defabg}
\end{equation}

We see that with the relations (\ref{sixcomms}), the six commutators with $J_i^{(1)}$ close on the following $36$-component multiplet of
obtained from six colour $\Gamma^{\mu}_{(a)}$ matrices:
\begin{equation}
{\cal{C}}^{+} = \left( \; \Gamma^0_{(1)}, \;  \Gamma^i_{(2)}, \; \Gamma^0_{(3)}, \;  \Gamma^i_{(4)}, \; \Gamma^0_{(5)}, \;  \Gamma^i_{(6)}, \; \right)
\label{Largegammaplus}
\end{equation} 
which describes the following triplet of $\Gamma^{\mu}$-matrices given by formula (\ref{GammaIexp}) with $\alpha=3$:
\begin{equation}
{\cal{C}}^{+} = \left( \Gamma^{\mu}_{(2;3)}, \; \Gamma^{\mu}_{(4;3)}, \; \Gamma^{\mu}_{(8;3)}, \right)
\label{Cplusthree}
\end{equation}
In a short-hand notation the relations (\ref{sixcomms}) look as follows:
\begin{equation}
[J^{(1)}_i, \Gamma_{(2;3)} ] \simeq \Gamma_{(4;3)}, \; \; \; \; [J^{(1)}_i, \Gamma_{(4;3)} ] \simeq \Gamma_{(8;3)}, \; \; \; \; 
[J^{(1)}_i, \Gamma_{(8;3)} ] \simeq \Gamma_{(2;3)},
\label{symbJ1}
\end{equation}
Now let us generate a new sequence of commutators starting from $\Gamma^k_{(2;2)}$:
$$\left[ J_i^{(1)}, \; \Gamma^k_{(2; 2)} \right] = - \frac{1}{2} \epsilon_{i \; \; \; m}^{\; \; k} \Gamma^m_{(4;2)} + 
\frac{\alpha}{2} \delta_i^k \Gamma^0_{(4;2)} = - \frac{1}{2} \epsilon_{i \; \; \; m}^{\; \; k} \Gamma^m_{(5)} + \frac{\alpha}{2} \delta_i^k \Gamma^0_{(6)};$$
$$\left[ J_i^{(1)}, \; \Gamma^0_{(4; 2)} \right] = - \frac{\gamma}{2} \Gamma^i_{(8; 2)} = - \frac{\gamma}{2} \Gamma^i_{(3)};$$
$$\left[ J_i^{(1)}, \; \Gamma^k_{(4; 2)} \right] = - \frac{j^2}{2} \epsilon_{i \; \; \; m}^{\; \; k} \Gamma^m_{(8;2)} + 
\frac{\gamma}{2} \delta_i^k \Gamma^0_{(8;2)} = - \frac{1}{2} \epsilon_{i \; \; \; m}^{\; \; k} \Gamma^m_{(3)} + \frac{\gamma}{2} \delta_i^k \Gamma^0_{(4)};$$
$$ \left[ J_i^{(1)}, \; \Gamma^0_{(8; 2)} \right] = - \frac{\gamma}{2} \Gamma^i_{(2;2)} = - \frac{\gamma}{2} \Gamma^i_{(1)}; $$
\begin{equation}
\left[ J_i^{(1)}, \; \Gamma^k_{(8; 2 )} \right] = -\frac{j}{2} \epsilon_{i \; \; \; m}^{\; \; k} \Gamma^m_{(2; 2)} + 
\frac{\beta}{2} \delta_i^k \; \Gamma^0_{(2;2)} = -\frac{j}{2} \epsilon_{i \; \; \; m}^{\; \; k} \Gamma^m_{(1)} + \frac{\beta}{2} \delta_i^k \; \Gamma^0_{(2)};
\label{sixcommsbis}
\end{equation}
$$\left[ J_i^{(1)}, \; \Gamma^0_{(2; 2)} \right] = -\frac{\alpha}{2}  \Gamma^i_{(4; 2)} = - \frac{\alpha}{2}  \Gamma^i_{(5)};$$
We get the $36$-component multiplet ${\cal{C}}^{-}$ of $12 \times 12$ dimensional matrices $\Gamma^{\mu}_{(a)}$:
\begin{equation}
{\cal{C}}^{-} = \left( \; \Gamma^0_{(2)}, \;  \Gamma^i_{(1)}, \; \Gamma^0_{(4)}, \;  \Gamma^i_{(3)}, \; \Gamma^0_{(6)}, \;  \Gamma^i_{(5)}, \; \right)
= \left( \Gamma^{\mu}_{(2;2)}, \; \Gamma^{\mu}_{(4;2)}, \; \Gamma^{\mu}_{(8;2)} \right),
\label{Largegammaminus}
\end{equation}
and the counterpart of the relations (\ref{symbJ1}) for the triplet ${\cal{C}}^{-}$ is obtained by replacing in the corresponding
$\Gamma$-matrices the indices $(A;3)$ by $(A;2)$.
The union ${\cal{C}}^{\mu} = \left( {\cal{C}}^{+} \oplus {\cal{C}}^{-} \right)$ describes the $72$-component {\it sextet} \;
${\boldsymbol{\Gamma}}^{\mu} = \left( \Gamma^{\mu}_{(1)}, ...., \Gamma^{\mu}_{(6)} \right)$ (see \ref{SextetG}),
and we obtain (\ref{Largegammaminus}) from (\ref{Largegammaplus}) by replacing in formula (\ref{GammaIexp}) $\sigma_{\alpha}=\sigma_3$
by  $\sigma_{\alpha}=\sigma_2$

\subsubsection{ Grade $1$ boosts generated by $K^{(1)}_i$ }
\vskip 0.2cm
By analogy with calculation of covariance under the generators $J_i^{(1)}$ we consider
now the closure of actions of the generators $K_i^{(1)}$ (see \ref{J2K1}) on the multiplets of colour $\Gamma^{\mu}$ matrices.
If we start with the boost transformations acting on the first standard colour matrices $\Gamma^{\mu}_{(1)} = (\Gamma^0_{(8 ;3)}, \Gamma^k_{(2;2)})$
of the sextet (\ref{Gammexpnum}), we obtain the closure after the calculation of the following set of $12$ commutators: 

$$ \left[ K_i^{(1)}, \; \Gamma^0_{(8; 3)} \right] = -\frac{j}{2} \Gamma^i_{(2; 2)} = -\frac{j}{2} \Gamma^i_{(1)} $$
$$ \left[ K_i^{(1)}, \; \Gamma^k_{(2; 2)} \right] = \frac{\alpha}{2} \epsilon^{\; \; k}_{i \; \; \; m} \Gamma^m_{(4;3)} 
- \frac{1}{2} \delta_i^k \; \Gamma^0_{(4;3)} = \frac{\alpha}{2} \epsilon^{\; \; k}_{i \; \; \; m} \Gamma^m_{(2)} 
- \frac{1}{2} \delta_i^k \; \Gamma^0_{(3)} =  $$
$$ \left[ K_i^{(1)}, \; \Gamma^0_{(4; 3)} \right] = -\frac{j^2}{2} \Gamma^i_{(8; 2)} = -\frac{j^2}{2} \Gamma^i_{(3)}$$ 
$$ \left[ K_i^{(1)}, \; \Gamma^k_{(4; 3)} \right] = -\frac{\gamma}{2} \epsilon^{\; \; k}_{i \; \; \; m} \Gamma^m_{(8;2)} 
+ \frac{j^2}{2} \delta_i^k \; \Gamma^0_{(8;2)}  = -\frac{\gamma}{2} \epsilon^{\; \; k}_{i \; \; \; m} \Gamma^m_{(3)}
+ \frac{j^2}{2} \delta_i^k \; \Gamma^0_{(4)} $$
$$ \left[ K_i^{(1)}, \; \Gamma^0_{(8; 2)} \right] = \frac{j}{2} \Gamma^i_{(2; 3)} = \frac{j}{2} \Gamma^i_{(4)} $$
$$ \left[ K_i^{(1)}, \; \Gamma^k_{(8; 2)} \right] = \frac{\beta}{2} \epsilon^{\; \; k}_{i \; \; \; m} \Gamma^m_{(2;3)}
- \frac{j}{2} \delta_i^k \; \Gamma^0_{(2;3)}  = \frac{\beta}{2} \epsilon^{\; \; k}_{i \; \; \; m} \Gamma^m_{(4)}
- \frac{j}{2} \delta_i^k \; \Gamma^0_{(5)} $$
\begin{equation}
 \left[ K_i^{(1)}, \; \Gamma^0_{(2; 3)} \right] = - \frac{1}{2} \Gamma^i_{(4; 2)} = - \frac{1}{2} \Gamma^i_{(5)}
\label{TwelveK1}
\end{equation}
$$ \left[ K_i^{(1)}, \; \Gamma^k_{(2; 3)} \right] = -\frac{\alpha}{2} \epsilon^{\; \; k}_{i \; \; \; m} \Gamma^m_{(4;2)} 
+ \frac{1}{2} \delta_i^k \; \Gamma^0_{(4;2)} = -\frac{\alpha}{2} \epsilon^{\; \; k}_{i \; \; \; m} \Gamma^m_{(5)} 
+ \frac{1}{2} \delta_i^k \; \Gamma^0_{(6)} = $$
$$ \left[ K_i^{(1)}, \; \Gamma^0_{(4; 2)} \right] = \frac{j^2}{2} \Gamma^i_{(8; 3)} =  \frac{j^2}{2} \Gamma^i_{(6)}  $$
$$ \left[ K_i^{(1)}, \; \Gamma^k_{(4; 2)} \right] = \frac{\gamma}{2} \epsilon^{\; \; k}_{i \; \; \; m} \Gamma^m_{(8;3)} 
- \frac{j^2}{2} \delta_i^k \; \Gamma^0_{(8;3)} =  \frac{\gamma}{2} \epsilon^{\; \; k}_{i \; \; \; m} \Gamma^m_{(6)} 
- \frac{j^2}{2} \delta_i^k \; \Gamma^0_{(1)} $$
$$ \left[ K_i^{(1)}, \; \Gamma^k_{(8; 3)} \right] = -\frac{\beta}{2} \epsilon^{\; \; k}_{i \; \; \; m} \Gamma^m_{(2;2)} 
- \frac{j}{2} \delta_i^k \; \Gamma^0_{(2;2)} = -\frac{\beta}{2} \epsilon^{\; \; k}_{i \; \; \; m} \Gamma^m_{(1)} 
- \frac{j}{2} \delta_i^k \; \Gamma^0_{(2)} $$
$$ \left[ K_i^{(1)}, \; \Gamma^0_{(2; 2)} \right] = \frac{1}{2} \Gamma^i_{(4; 3)} = \frac{1}{2} \Gamma^i_{(2)}  $$
The relations (\ref{TwelveK1}) can be also expressed with a short-hand notation as follows
\begin{equation} 
[K^{(1)}, C^{+} ] \simeq C^{-}, \; \; \; \; [K^{(1)}, C^{-} ] \simeq C^{+}.
\label{symbK1}
\end{equation} 

We see that all $72$ components of the sextet (\ref{SextetG}) are needed in order to obtain the irreducible representation 
closed under the action of the boost generators $K^{(1)}_i$.

The pattern of the coefficients appearing on the right-hand side of these $12$ commutators bears the imprint of the underlying $Z_3 \times Z_2$ symmetry.
The six commutators of $K^{(1)}_i$ with time-like components of $\Gamma$-matrices produce only the space-like components, multiplied by
 halves of all sixth-order roots of unity, i.e. $\pm \frac{1}{2}, \; \pm \frac{j}{2}, \; \pm \frac{j^2}{2}$, while the commutators with spatial
components $\Gamma^k_{(A; \alpha)}$ contain again the spatial components, multiplied by the coefficients 
$\pm \frac{\alpha}{2}, \pm \frac{\beta}{2}, \pm \frac{\gamma}{2}$ and time-like components $\Gamma^0_{(A; \alpha)}$ with the coefficients 
$\pm \frac{1}{2}, \; \pm \frac{j}{2}, \; \pm \frac{j^2}{2}$, 
The full multiplication table of this Lie algebra over complex roots together with the diagram
showing the structure constants on the complex plane are given in the Appendix I.

It is worth to observe that in the definitions (\ref{sixGammas}) of the basic sextet ${\boldsymbol{\Gamma}}^{\mu}$ in the colour sector enters only
 the following {\it triplet} of colour generators
$(I_2, \; I_4, \; I_8) = ( Q_2, \;  Q^{\dagger}_1, \; B^{\dagger})$ which satisfies the following relations 
(see also the Table of commutators in Appendix I):
\begin{equation}
\left[ Q_3, \; Q_2 \right] = \alpha \; Q^{\dagger}_1, \; \; \left[ Q_3, \; Q_1^{\dagger} \right] = \beta \; B^{\dagger}, \; \; 
\left[ Q_3, \; B^{\dagger} \right] = \gamma \; Q_2, 
\label{commwithQ3}
\end{equation}
The closure of the action of $Q_3$ on the multiplet $(Q_2, \; Q^{\dagger}_1, \; B^{\dagger})$ leads to the covariance of the 
$72$-dimensional multiplet (\ref{sixGammasdag}) under the action of the generators $\left( J^{(1)}_i, \; K^{(1)}_m \right) \in L^{(1)}$, 
which do contain the matrix $Q_3$ as their first colour factor (see (\ref{J2K1}, \ref{J1K2})).

It can be recalled (see \ref{sixGammasdag}) that in order to construct the Lorentz doublets $(\Gamma_{(a)}^{\mu}, \; {\tilde{\Gamma}}_{(a)}^{\mu})$
$(a=1,2,...,6)$ it is sufficient to use the components of the sextet of matrices $\Gamma_{(a)}^{\mu}$ suffice (see (\ref{SextetG}), and again the relations (\ref{commwithQ3}) 
imply the closure of ${\tilde{\Gamma}}^{\mu}_{(a)}$ under the actions of generators belonging to $L^{(1)}$. 

\subsection{ Representations of $Z_3$-graded Lorentz algebra - sector $L^{(2)} = J^{(2)}_i \oplus K^{(2)}_m $ }
\vskip 0.2cm
\indent
The matrices of the sextet multiplet ${\boldsymbol{\Gamma}}^{\mu}$ are complex and non-Hermitean. Due to the relations
\begin{equation}
\left( J_i^{(1)} \right)^{\dagger} = - J_i^{(2)}, \; \; \; \left( K_i^{(1)} \right)^{\dagger} =  K_i^{(2)}, 
\label{HermitJK}
\end{equation}
in order to obtain the closed action of grade $2$ generators one should introduce 
the Hermitean-conjugate sextet ${\boldsymbol{\Gamma}}^{\dagger}$ of $\Gamma^{\mu}$-matrices (\ref{sixGammas})
\begin{equation}
\left( {\boldsymbol{\Gamma}}^{\mu} \right)^{\dagger} = \left( \left(\Gamma^{\mu}_{(1)} \right)^{\dagger}, \; 
...., \left(\Gamma^{\mu}_{(6)} \right)^{\dagger} \right)
\label{HermGammas}
\end{equation}  
One can deduce from the relations (\ref{sixcomms}), (\ref{TwelveK1}) and (\ref{HermitJK}) the covariant actions of generators 
from the sector $L^{(2)}$ by using the formulae
{\small
\begin{equation}
\left( \left[ J_i^{(1)}, \; \Gamma^{\mu}_{(A; \alpha)} \right] \right)^{\dagger} = 
- \left[ \left( J_i^{(1)} \right)^{\dagger}, \; \left( \Gamma^{\mu}_{(A; \alpha)} \right)^{\dagger} \right]
=  \left[ J_i^{(2)} , \; \left( \Gamma^{\mu}_{(A; \alpha)} \right)^{\dagger} \right],
\label{J12perm}
\end{equation}
\begin{equation}
\left( \left[ J_i^{(2)}, \; \Gamma^{\mu}_{(A; \alpha)} \right] \right)^{\dagger} = 
- \left[ \left( J_i^{(2)} \right)^{\dagger}, \; \left( \Gamma^{\mu}_{(A; \alpha)} \right)^{\dagger} \right]
=  \left[ J_i^{(1)} , \; \left( \Gamma^{\mu}_{(A; \alpha)} \right)^{\dagger} \right],
\label{J21perm}
\end{equation}
\begin{equation}
\left( \left[ K_i^{(1)}, \; \Gamma^{\mu}_{(A; \alpha)} \right] \right)^{\dagger} = 
-  \left[ \left( K_i^{(1)} \right)^{\dagger}, \; \left( \Gamma^{\mu}_{(A; \alpha)} \right)^{\dagger} \right]
= -  \left[  K_i^{(2)} , \; \left( \Gamma^{\mu}_{(A; \alpha)} \right)^{\dagger} \right],
\label{K12perm}
\end{equation}
\begin{equation}
\left( \left[ K_i^{(2)}, \; \Gamma^{\mu}_{(A; \alpha)} \right] \right)^{\dagger} = 
-  \left[ \left( K_i^{(2)} \right)^{\dagger}, \; \left( \Gamma^{\mu}_{(A; \alpha)} \right)^{\dagger} \right]
= -  \left[  K_i^{(1)} , \; \left( \Gamma^{\mu}_{(A; \alpha)} \right)^{\dagger} \right].
\label{K21perm}
\end{equation} }

One obtains in such a way the irreducible actions of sector $L^{(2)}$ on the sextet (\ref{HermGammas}) 
of Hermitean-conjugated $\Gamma^{\mu}$-matrices which describe the grade $2$ counterparts of the relations (\ref{sixcomms}), 
(\ref{sixcommsbis}) and (\ref{TwelveK1}). In the sector $L^{(2)}$ one can introduce as well the Hermitean-conjugate  Lorentz doublets
$\left( (\Gamma_{(a)}^{\mu})^{\dagger}, \; ({\tilde{\Gamma}}_{(a)}^{\mu}))^{\dagger} \right)$ 

The commutators of the sextet ${\boldsymbol{\Gamma}}$ (see (\ref{sixGammasdag})) with grade $2$ generators $J^{(2)}_i$ of spatial rotations form
the same two closed sets of relations for the multiplets (\ref{Largegammaplus}) and (\ref{Largegammaminus}) (see also formula (\ref{GammaIexp}),
which are analogous to the realizations of the generators $J^{(1)}_i$, with grade $1$. More explicitely,

$$ \left[ J_i^{(2)}, \; \Gamma^0_{(8; 3)} \right] = -\frac{\beta}{2} \Gamma^i_{(4; 3)} = -\frac{\beta}{2} {\tilde{\Gamma}}^i_{(6)};$$
$$ \left[ J_i^{(2)}, \; \Gamma^k_{(4; 3)} \right] = - \frac{1}{2} \epsilon^{\; \; k}_{i \; \; \; m} \Gamma^m_{(2;3)} 
- \frac{\alpha}{2} \delta_i^k \; \Gamma^0_{(2;3)}
= - \frac{1}{2} \epsilon^{\; \; k}_{i \; \; \; m} {\tilde{\Gamma}}^m_{(2)} 
- \frac{\alpha}{2} \delta_i^k \; {\tilde{\Gamma}}^0_{(1)}$$
$$ \left[ J_i^{(2)}, \; \Gamma^0_{(2; 3)} \right] = \frac{\gamma}{2} \Gamma^i_{(8; 3)} = \frac{\gamma}{2} {\tilde{\Gamma}}^i_{(4)}$$
\begin{equation}
 \left[ J_i^{(2)}, \; \Gamma^k_{(2; 3)} \right] = -\frac{j}{2} \epsilon^{\; \; k}_{i \; \; \; m} \Gamma^m_{(8;3)} - 
\frac{\beta}{2} \delta_i^k \; \Gamma^0_{(8;3)}
= - \frac{j}{2} \epsilon^{\; \; k}_{i \; \; \; m} {\tilde{\Gamma}}^m_{(4)} - \frac{\beta}{2} \delta_i^k \; {\tilde{\Gamma}}^0_{(3)} 
\label{J2commfirst}
\end{equation}
$$ \left[ J_i^{(2)}, \; \Gamma^0_{(4; 3)} \right] = - \frac{\alpha}{2} \Gamma^i_{(2; 3)} = - \frac{\alpha}{2} {\tilde{\Gamma}}^i_{(2)} $$
$$ \left[ J_i^{(2)}, \; \Gamma^k_{(8; 3)} \right] = - \frac{j}{2} \epsilon^{\; \; k}_{i \; \; \; m} \Gamma^m_{(4;3)} 
+ \frac{\beta}{2} \delta_i^k \; \Gamma^0_{(4;3)}
=   - \frac{j}{2} \epsilon^{\; \; k}_{i \; \; \; m} {\tilde{\Gamma}}^m_{(6)} + \frac{\beta}{2} \delta_i^k \; {\tilde{\Gamma}}^0_{(5)} $$
The next chain of commutators with $J^{(2)}$ we begin with $\Gamma^k_{(2;2)}$:

$$ \left[ J_i^{(2)}, \; \Gamma^k_{(2; 2)} \right] = - \frac{j^2}{2} \epsilon^{\; \; k}_{i \; \; \; m} \Gamma^m_{(8;2)} 
- \frac{\gamma}{2} \delta_i^k \; \Gamma^0_{(8;2)} =
- \frac{j^2}{2} \epsilon^{\; \; k}_{i \; \; \; m} {\tilde{\Gamma}}^m_{(1)} - \frac{\gamma}{2} \delta_i^k \; {\tilde{\Gamma}}^0_{(6 )}$$
$$ \left[ J_i^{(2)}, \; \Gamma^0_{(8; 2)} \right] = \frac{\beta}{2} \Gamma^i_{(4; 2)} = = \frac{\beta}{2} {\tilde{\Gamma}}^i_{(3)}$$ 
$$ \left[ J_i^{(2)}, \; \Gamma^k_{(8; 2)} \right] = -\frac{j}{2} \epsilon^{\; \; k}_{i \; \; \; m} \Gamma^m_{(4;2)} 
- \frac{\beta}{2} \delta_i^k \; \Gamma^0_{(4;2)} =
-\frac{j}{2} \epsilon^{\; \; k}_{i \; \; \; m} {\tilde{\Gamma}}^m_{(3)} - \frac{\beta}{2} \delta_i^k \; {\tilde{\Gamma}}^0_{(2)}$$
\begin{equation}
\left[ J_i^{(2)}, \; \Gamma^0_{(4; 2)} \right] = - \frac{\alpha}{2} \Gamma^i_{(2; 2)} = - \frac{\alpha}{2} {\tilde{\Gamma}}^i_{(5)}
\label{J2commsecond}
\end{equation}
$$\left[ J_i^{(2)}, \; \Gamma^k_{(4; 2)} \right] = - \frac{1}{2} \epsilon^{\; \; k}_{i \; \; \; m} \Gamma^m_{(2;2)} 
+ \frac{\alpha}{2} \delta_i^k \; \Gamma^0_{(2;2)}
=- \frac{1}{2} \epsilon^{\; \; k}_{i \; \; \; m} {\tilde{\Gamma}}^m_{(5)} + \frac{\alpha}{2} \delta_i^k \; {\tilde{\Gamma}}^0_{(4)} $$
$$ \left[ J_i^{(2)}, \; \Gamma^0_{(2; 2)} \right] = \frac{\gamma}{2} \Gamma^i_{(8; 2)} = \frac{\gamma}{2} {\tilde{\Gamma}}^i_{(1)} $$

Similar series of commutators, starting with the ``basic'' colour Dirac matrix $\Gamma^{\mu}_{(1)}$ and then continuing to closed
commutators structure is produced by the following actions of boosts from the grade $2$ Lorentz sector $L^{(2)}$:
 
$$ \left[ K_i^{(2)}, \; \Gamma^0_{(8; 3)} \right] = -\frac{j}{2} \Gamma^i_{(4; 2)} = -\frac{j}{2} {\tilde{\Gamma}}^i_{(3)};$$
$$ \left[ K_i^{(2)}, \; \Gamma^k_{(2; 2)} \right] = - \frac{\gamma}{2} \epsilon^{\; \; k}_{i \; \; \; m} \Gamma^m_{(8;3)} 
- \frac{j^2}{2} \delta_i^k \; \Gamma^0_{(8;3)}
= - \frac{\gamma}{2} \epsilon^{\; \; k}_{i \; \; \; m} {\tilde{\Gamma}}^m_{(4)} - \frac{j^2}{2} \delta_i^k \; {\tilde{\Gamma}}^0_{(3)}   $$
$$ \left[ K_i^{(2)}, \; \Gamma^0_{(2; 2)} \right] = \frac{j^2}{2} \Gamma^i_{(8; 3)} = \frac{j^2}{2} \Gamma^i_{(4)} $$ 
$$ \left[ K_i^{(2)}, \; \Gamma^k_{(8; 3)} \right] = \frac{\beta}{2} \epsilon^{\; \; k}_{i \; \; \; m} \Gamma^m_{(4;2)} 
+ \frac{j}{2} \delta_i^k \; \Gamma^0_{(4;2)}
= \frac{\beta}{2} \epsilon^{\; \; k}_{i \; \; \; m} {\tilde{\Gamma}}^m_{(3)} + \frac{j}{2} \delta_i^k \; {\tilde{\Gamma}}^0_{(2)} $$
$$ \left[ K_i^{(2)}, \; \Gamma^0_{(2; 3)} \right] = - \frac{j}{2} \Gamma^i_{(8; 2)} = - \frac{j}{2} {\tilde{\Gamma}}^i_{(1)} $$
\begin{equation}
\left[ K_i^{(2)}, \; \Gamma^k_{(4; 2)} \right] = - \frac{\alpha}{2} \epsilon^{\; \; k}_{i \; \; \; m} \Gamma^m_{(2;3)} 
- \frac{1}{2} \delta_i^k \; \Gamma^0_{(2;3)}
= - \frac{\alpha}{2} \epsilon^{\; \; k}_{i \; \; \; m} {\tilde{\Gamma}}^m_{(2)} - \frac{1}{2} \delta_i^k \; {\tilde{\Gamma}}^0_{(1)}
\label{AllcommK2}
\end{equation}
$$ \left[ K_i^{(2)}, \; \Gamma^0_{(4; 2)} \right] = \frac{1}{2} \Gamma^i_{(2; 3)} =  \frac{1}{2} \Gamma^i_{(2)} $$
$$ \left[ K_i^{(2)}, \; \Gamma^k_{(2; 3)} \right] = \frac{\gamma}{2} \epsilon^{\; \; k}_{i \; \; \; m} \Gamma^m_{(8;2)} 
+ \frac{j^2}{2} \delta_i^k \; \Gamma^0_{(8;2)}
= \frac{\gamma}{2} \epsilon^{\; \; k}_{i \; \; \; m} {\tilde{\Gamma}}^m_{(1)} + \frac{j^2}{2} \delta_i^k \; {\tilde{\Gamma}}^0_{(6)}  $$
$$ \left[ K_i^{(2)}, \; \Gamma^0_{(8; 2)} \right] = \frac{j}{2} \Gamma^i_{(4; 3)} = \frac{j}{2} {\tilde{\Gamma}}^i_{(6)} $$
$$ \left[ K_i^{(2)}, \; \Gamma^k_{(8; 2)} \right] = - \frac{\beta}{2} \epsilon^{\; \; k}_{i \; \; \; m} \Gamma^m_{(4;3)} 
- \frac{j}{2} \delta_i^k \; \Gamma^0_{(4;3)}
= - \frac{\beta}{2} \epsilon^{\; \; k}_{i \; \; \; m} {\tilde{\Gamma}}^m_{(6)} - \frac{j}{2} \delta_i^k \; {\tilde{\Gamma}}^0_{(5)} $$
$$ \left[ K_i^{(2)}, \; \Gamma^0_{(4; 3)} \right] = - \frac{1}{2} \Gamma^i_{(2; 2)} = - \frac{1}{2} {\tilde{\Gamma}}^i_{(5)}  $$
$$ \left[ K_i^{(2)}, \; \Gamma^k_{(4; 3)} \right] = \frac{\alpha}{2} \epsilon^{\; \; k}_{i \; \; \; m} \Gamma^m_{(2;2)} 
+ \frac{1}{2} \delta_i^k \; \Gamma^0_{(2;2)}
= \frac{\alpha}{2} \epsilon^{\; \; k}_{i \; \; \; m} {\tilde{\Gamma}}^m_{(2;2)} + \frac{1}{2} \delta_i^k \; {\tilde{\Gamma}}^0_{(2;2)}  $$

Because from (\ref{commwithQ3}) follows the closure of the triplet 
$$(I_1, \; I_5, \; I_7) \equiv (I_4^{\dagger}, I^{\dagger}_2, \;  \; I^{\dagger}_8)) = (Q_1, \; Q^{\dagger}_2, \; B),$$
under the action of $Q^{\dagger}_3$, (compare with (\ref{IAdef})) one can as well reproduce the covariant action of $L^{(2)}$
on the Hermitean-conjugate doublets of $\Gamma^{\mu}$-matrices. 

The general pattern of commutators in (\ref{J2commfirst})-(\ref{AllcommK2}) better explains why the irreducible representation is described by
the sextet of colour $\Gamma^{\mu}$ matrices. The generators $J^{(1)}_i, J^{(2)}_i, K^{(1)}_m,  K^{(2)}_m$ contain as their $3 \times 3$ matrix
factors the elements $Q_3$ and $Q^{\dagger}_3$, which therefore cannot appear in the coloured $\Gamma^{\mu}$-matrices; the boosts contain 
as their second factor 
the matrix $\sigma_1$, which as well can not appear in the sextet (\ref{sixGammas}). Starting from the first ``standard'' colour Dirac operator 
whose $\Gamma$-matrices
contain $B^{\dagger}$ and $Q_2$, commutators with $Q_3$ and $Q^{\dagger}_3$ can generate in the colour sector only the third colour matrix
$Q^{\dagger}_1 $, besides $B^{\dagger}$ and $Q_2$. This reduces the number of $\Gamma^{\mu}$ matrices spanning the spinorial realization 
of the $Z_3$-graded Lorentz algebra to six, characterized by three colour matrices $I_A$ ($A = 2, \;  4, \; 8$) and two Pauli matrices 
$\sigma_{\alpha}$ ($\alpha = 2, \; 3$).

If we start with complex conjugate Dirac operator (\ref{Gammasecond}), (\ref{Gammasbig}) with $\Gamma^0 = B \otimes \sigma_3 \otimes {\mbox{l\hspace{-0.55em}1}}_{2}$ and
$\Gamma^i = Q_1 \otimes (i \sigma_2) \otimes \sigma^i$ (note that $B$ is the complex conjugate of $B^{\dagger}$ and $Q_1$ is the complex conjugate
of $Q_2$), we get the alternative sextet 
describing coloured Dirac equations for the complex-conjugated fields $\bar{\Psi}$ (see (\ref{PsiPsibar})), which contains as its colour factors 
the matrices $Q_2^{\dagger}, Q_1$ and $B$.

\section{Irreducible realizations of the $Z_3$-graded Lorentz algebra and the full set of quark symmetries}
\vskip 0.2cm
\subsection{Chiral colour doublets and flavour states}
\vskip 0.2cm
The flavour quark eigenstates in the Standard Model are represented by chirally projected Dirac spinors (see e.g. \cite{Sogami2019}, \cite{Sogami2020}).
If we introduce the $D=3+1$ Clifford algebra defined by the relations
\begin{equation}
\{ \gamma^{\mu}, \; \gamma^{\nu} \} = 2 \; \eta^{\mu \nu} \;  {\mbox{l\hspace{-0.55em}1}}_{4}, \; \; \; 
\eta^{\mu \nu} = {\rm diag} (1, -1, -1, -1),
\label{Clifford4}
\end{equation}
and define 
\begin{equation}
\gamma^5 = \gamma^0 \gamma^1 \gamma^2 \gamma^3, \; \; \; (\gamma^5)^2 = - {\mbox{l\hspace{-0.55em}1}}_{4}
\label{gamma5}
\end{equation}
then the standard chiral Dirac spinors are defined as
\begin{equation}
\psi_{\pm} = \pm i \gamma^5 \; \psi_{\pm} = \frac{1}{2} (1 \pm i \gamma^5) \psi.
\label{chirals}
\end{equation}
The  $\psi_{\pm}$ denote the four-component Dirac spinors satisfying the chirality conditions
$P_{\pm} \; \psi_{\pm} = \psi_{\pm}, \; \; P_{\mp} \psi_{\pm} = 0$, where  $P_{\pm}= \frac{1}{2} ({\mbox{l\hspace{-0.55em}1}}_{2} \pm i \gamma^5)$
are the chiral projection operators.

For more clarity we shall use the following realization of Clifford algebra of Dirac matrices in terms
of tensor products of $2 \times 2$ matrices:
\begin{equation}
\gamma^0 = \sigma_3 \otimes \; {\mbox{l\hspace{-0.55em}1}}_{2}, \; \; \; 
\gamma^i = (i \sigma_2) \otimes \sigma^i, \; \; \; \; \gamma^5 =- i \sigma_1 \otimes \; {\mbox{l\hspace{-0.55em}1}}_{2}.
\label{gammatensor}
\end{equation}
This enables us to express the colour Dirac $12 \times 12$-matrices in a more concise manner:
$$\Gamma^0_{(1)} = \Gamma^0_{(8,3)} = B^{\dagger} \otimes \gamma^0, \; \; \; \; 
\Gamma_{(1)}^i =  \Gamma^0_{(2,2)} = Q_2 \otimes \gamma^i,$$
\begin{equation} 
{\tilde{\Gamma}}^0_{(1)} = \Gamma^0_{(2,3)} = Q_2 \otimes \gamma^0, \; \; \; \; 
{\tilde{\Gamma}}_{(1)}^i =  \Gamma^0_{(8,2)} = B^{\dagger} \otimes \gamma^i.
\label{gammasshort}
\end{equation}
The chiral projection operator acting on the matrices $\Gamma_{(1)}^{\mu}$ can be now defined as follows:
\begin{equation}
P_{\pm} = {\mbox{l\hspace{-0.55em}1}}_{3} \otimes \frac{1}{2}({\mbox{l\hspace{-0.55em}1}}_{2} \pm \sigma_1) 
\otimes {\mbox{l\hspace{-0.55em}1}}_{2},
\label{projpm}
\end{equation}
so that the chirally projected matrices $\Gamma^{\mu}_{(1) \pm} = P_{\pm} \Gamma^{\mu}_{(1)}$ look as follows:
$$P_{\pm} \Gamma_{(1)}^{\mu} = \Gamma^{\mu}_{(1) \pm} = 
\left( B^{\dagger} \otimes \frac{1}{2} (\sigma_3 \mp i \sigma_2) \otimes \; 
{\mbox{l\hspace{-0.55em}1}}_{2}, \; \; Q_2 \otimes \frac{1}{2} (i \sigma_2 \mp \sigma_3) 
\otimes \sigma^i \; \right) =$$
\begin{equation}
= \left( \frac{1}{2} (\Gamma^0_{(1)} \mp i \Gamma^0_{(4)}) , \; \; 
\frac{1}{2} (\Gamma^i_{(1)} \pm \Gamma^i_{(4)}) \right)
\label{chralgammasbig1}
\end{equation} 
By adding the relations

$$P_{\pm} \Gamma_{(2)}^{\mu} = \Gamma^{\mu}_{(2) \pm} = 
\left( Q_2 \otimes \frac{1}{2} (\sigma_2 \pm \sigma_3) \otimes \; 
{\mbox{l\hspace{-0.55em}1}}_{2}, \; \; Q_1^{\dagger} \otimes \frac{1}{2} (\sigma_3 \pm i \sigma_2) 
\otimes \sigma^i \; \right) =$$
\begin{equation}
= \left( \frac{1}{2} (\Gamma^0_{(2)} \pm \Gamma^0_{(5)}) , \; \; 
\frac{1}{2} (\Gamma^i_{(2)} \pm \Gamma^i_{(5)}\right)
\label{chralgammasbig2}
\end{equation}

$$P_{\pm} \Gamma_{(3)}^{\mu} = \Gamma^{\mu}_{(3) \pm} = 
\left( Q_1^{\dagger} \otimes \frac{1}{2} (\sigma_3 \pm \sigma_2) \otimes \; 
{\mbox{l\hspace{-0.55em}1}}_{2}, \; \; B^{\dagger} \otimes \frac{1}{2} (i \sigma_2 \pm \sigma_3) 
\otimes \sigma^i \; \right) =$$
\begin{equation}
= \left( \frac{1}{2} (\Gamma^0_{(3)} \pm \Gamma^0_{(6)}) , \; \; 
\frac{1}{2} (\Gamma^i_{(1)} \pm \Gamma^i_{(4)}\right)
\label{chralgammasbig3}
\end{equation} 
one introduces chiral/anti-chiral triplets $\Gamma^{\mu}_{(r)} \; (r = 1,2,3)$ of colour $\Gamma$-matrices defined in terms of three pairs 
$(\Gamma^{\mu}_{(1)}, \Gamma^{\mu}_{(4)}),$
$(\Gamma^{\mu}_{(2)}, \Gamma^{\mu}_{(5)}),$ $(\Gamma^{\mu}_{(3)}, \Gamma^{\mu}_{(6)}),$
or six matrices $\Gamma^{\mu}_{(A; \alpha)}$ defined by formula (\ref{GammaIexp}).
It is interesting to note that one can construct similar projection operators related with $Z_3$-graded colour sectors. Let us define
three projection operators:
\begin{equation}
{\overset{(0)}{\Pi}} = \frac{1}{3}  ({\mbox{l\hspace{-0.55em}1}}_{3} + {\cal{B}} + {\cal{B}}^{\dagger}) \otimes \; {\mbox{l\hspace{-0.55em}1}}_{4}, \; \; \; 
\label{threeproj0}
\end{equation}
\begin{equation}
{\overset{(1)}{\Pi}} = \frac{1}{3} ({\mbox{l\hspace{-0.55em}1}}_{3} + j^2 \; {\cal{B}} + j \; {\cal{B}}^{\dagger}) \otimes \; {\mbox{l\hspace{-0.55em}1}}_{4}, \; \; \; 
{\overset{(2)}{\Pi}} = \frac{1}{3} ({\mbox{l\hspace{-0.55em}1}}_{3} + j \; {\cal{B}} + j^2 \; {\cal{B}}^{\dagger}) \otimes \; {\mbox{l\hspace{-0.55em}1}}_{4}.
\label{threeproj}
\end{equation}
One checks easily that the three projectors (\ref{threeproj0}, \ref{threeproj}) satisfy the expected relations
\begin{equation}
[{\overset{(r)}{\Pi}}]^2 = {\overset{(r)}{\Pi}}, \; r=0,1,2, \; \; {\overset{(r)}{\Pi}} {\overset{(s)}{\Pi}} =0, \; \; {\rm for} \; \; r \neq s,
\; \; \; {\overset{(0)}{\Pi}} + {\overset{(1)}{\Pi}} + {\overset{(2)}{\Pi}} =  \; {\mbox{l\hspace{-0.55em}1}}_{3},  
\label{PiorthoB}
\end{equation}
i.e. we obtain a $Z_3$-graded generalization of the $Z_2$-graded standard chiral projectors $P_{\pm}$, where $P^2_{+} = P_{+}, \; \; 
P^2_{-} = P_{-}$, $P_{+} P_{-} = P_{-} P_{+} = 0$ and $P_{+}+ P_{-} = {\mbox{l\hspace{-0.55em}1}}.$  

\vskip 0.3cm
\subsection{The flavour and generations in quark models and chiral colour Dirac multiplets}
\vskip 0.2cm
Three generations of quarks (called also ``three families'') are known, each formed by a flavour (``weak isospin'') doublet:
\vskip 0.2cm
$\bullet$ \hskip 0.3cm $(u, \; d)$, \; or  the  ``up - down'' doublet (``First generation'');
\vskip 0.2cm 
$\bullet$ \hskip 0.3cm $(s, \; c)$, \; or  the  ``strange - charm'' doublet (``Second generation'');
\vskip 0.2cm 
$\bullet$ \hskip 0.3cm $(t, \; b)$, \; or  the  ``top - bottom'' doublet (``Third generation'');
\vskip 0.2cm 
The flavour $SU(2)$ symmetry is visible only if we consider chiral (left-handed) quarks, described by the doublet $(u, \; d)$.
In the case of $2^{\rm nd}$ and $3^{\rm rd}$ generation, possible flavour symmetries exchanging $c$ with $s$ or
$t$ with $b$ are strongly violated and the internal symmetry is used mostly for the classification purposes.
\footnote{Historically, for the classification purposes, the flavours were firstly ordered into the $SU(3)$ $(u, \; d, \; s)$ multiplet 
however at present we know that dynamically the coset $SU(3)/U(2)$ is badly broken}

It is well established that the first generation splits into an $SU(2)$ chiral flavour doublet $u_{+}, \; d_{+}$ and a pair
of anti-chiral flavour singlets $u_{-}$ and $d_{-}$. In order to introduce three quark generations as described by
second (non-colour) $SU(3)$ internal symmetry, one should also assume analogous chiral structures for the doublets $(s, \; c)$ and $(t, \; b)$. 
However, in phenomenological Lagrangeans, if we take into consideration quark interactions without imposing a priori chiral structure of Feynman 
graph vortices, it appears quite reasonable to keep the doublets $(s, \; c)$ and $(t, \; b)$ as non-chiral ones (see e.g. \cite{Sogami2020}) 

Having introduced chiral and anti-chiral colour Dirac matrices, we can define respective $12$-component chiral colour Dirac spinors and corresponding 
chiral colour Dirac equations (see (\ref{systemsix})).
The chiral and anti-chiral states  can be formed by pairs of quarks $(s, \; c)$ and $(t, \; b)$ of other generations, i.e. one can use 
the same scheme for the doublets in each of three generations.  

The chiral structure of the flavour sector of the Standard Model becomes important when we consider together leptons
and quarks, with leptons interacting weakly as a kind of fourth colour (see e.g. ({\cite{PatiSalam})). In our model one
can introduce leptons as colourless quarks just by replacing the $3 \times 2$ matricees $B$ and $Q_a$ appearing in the
tensor products defining the generalized $12 \times 12$ Dirac matices by the $3 \times 3$ unit matrix.

In the Standard Model the fact that leptons and coloured quarks are coupled weakly in analogous way leads to 
an imporant feature of the chiral anomaly cancellation. We hope that in the next stage of development of our model
with interaction vertices introduced, such a cancellation mechanism can be also naturally achieved.

\section{Outlook}
\vskip 0.2cm

The Standard Model (SM) of elementary particles is without doubt very successful experimentally tested part of theoretical physics; however,
its group-theoretical structure still requires further investigations. The internal symmetries are the product of three unitary groups
$SU(3) \times SU(2) \times U(1)$, with chiral $SU(2)$ sector describing weak interactions and basic role of colour $SU(3)$ 
group describing strongly interacting gluons and quarks which are not observable as free 
asymptotic states. The full spectrum of quarks requires still another $SU(3)$ symmetry due to appearance of quark generations,
which should be interpreted with the help of an additional geometric structure, describing from group-theoretical point 
of view the full set of all quarks as given by irreducible $72$-dimensional representations of a new group which
intertwines Lorentz and colour symmetries.

Unifying efforts in the literature ((see e.g. \cite{Sogami2019}, \cite{MarschNarita}, \cite{Sitarz}, 
\cite{Todorov}, \cite{Bochniak}) went along various paths, with two basic ways 
of unification: the first preserves the tensor product structure of space-time and
 internal symmetries, while the other one is more radical, intertwining 
the relativistic and internal colour symmetries. A well-known example of secon type of unification scheme is provided by the known passage 
from bosonic symmetries to supersymmetries, which describe the supermultiplet containing commuting bosons and anti-commuting fermions 
both incorporated in one common $Z_2$-graded algebraic structure. 

In our paper we deal exclusively with fundamental quark degrees of freedom, described by the collection of anti-commuting fields, with triplets of 
quark states with three different colours, which permit the introduction of $Z_3$-graded algebraic structure. The $Z_3$-grading does not change
the fermionic statistics of quarks, but leads to particular link between relativistic (Lorentz) symmetry and internal (colour) symmetries,
which in colour Dirac equations cease to be described by a tensor product group structure. The colour
generators are naturally expressed in a $Z_3$-graded ternary basis (see Sect. 2) and the $SU(3)$ colour
symmetries provide all possible $3 \times 3$ matrix choices of ternary basis (see the end of Sect. 4).
It is the $Z_3$-covariance which in a field-theoretic description of quarks provides the passage from 
the three copies of $4$-component standard Dirac fields 
to the $12$-dimensional colour Dirac field (see Sect. $3$). 

The physical observation that the quarks are described by six colour triplets did lead us to 
the idea that one should look for
an algebraic scheme which would provide a unifying $72$-dimensional module incorporating all six colour triplets of fermions inside 
a single irreducible representation. We demonstrate in this paper how this goal can be achieved by introduction 
of the $Z_3$-graded Lorentz symmetries, which can be extended to $Z_3$-graded Poincar\'e group. We consider the vectorial representation
of the $Z_3$-graded Poincar\'e algebra in Sect. $4$, and the spinorial representation of the $Z_3$-graded Lorentz algebra in Sect. $5$ and $6$. 

Usually the efforts to incorporate all existing quarks in a unique irreducible multiplet are restricted only to the discussion of 
internal degrees of freedom. Our approach, which leads to a $Z_3$-graded extension of standard relativistic symmetries, 
implies as well the modification of quark dynamics, what can be seen already from the wave equations for free quarks 
which imply dispersion relations of the sixth order, 
satisfied by all the components of the sextet of free $Z_3$-graded quark fields. The dispersion relations can be described as a triple
product of mass shells, one with real mass and a pair of mutually conjugate complex ones, 
which lead to the appearance of complex wave vectors (see Sect. $3$ and $4$) and provide damped exponential solutions along with
freely propagating waves. 

%Such modification, as we suppose, can be related with the confinement of asymptotic quark fields.
% already on the kinematical level.

Our model is formulated only on the preliminary kinematic level, without defining neither the Lagrangean, nor the
interaction vertices. Still on the kinematic level it is important to describe besides quarks, also the leptons
and gauge fields. Before we proceed further we observe that:

\indent
\hskip 0.3cm
i) One should complete the quantum field-theoretic description of free quantum $Z_3$-graded fermionic quark fields,
with algebra of field oscillators and Green function.

In this paper we provided only the formulae for the quark propagators (see (\ref{Dirac3inverse}), (\ref{InverseK}),
(\ref{InverseK2})). These formulae follow from the decomposition of basic quantum quark field satisfying sixth order 
equation into three Klein-Gordon like fields with $Z_3$-graded set of complex masses $m_s = j^s m, \; (s = 0,1,2)$.
Further one should consider the $Z_3$-covariant set of quantized free Klein-Gordon fields with respective oscillator 
algebras describing the field quanta that lead to three residua $(1, j, j^2 )$ of three propagators in eq. (\ref{InverseK}).  
These residua describe three respective metrics in the Hilbert-Fock spaces associated with our three Klein-Gordon type free quantum fields.
\indent
\hskip 0.3cm
ii) One should provide the $Z_3$-covariant interaction vertices, in particular the prescriptions for gauge field couplings.
In order to obtain such results we should study the role of $Z_3$-graded Lorentz transformations in space-time
( Sect. 3 and 4 were restricted only to the four-momentum space).

In order to be able to construct the action density and covariantize the space-time
derivatives by introducing gauge fields we should find out how the $Z_3$-graded Lorentz and Poincar\'e generators act
on the space-time coordinates. This is going to be the subject of our future research.

%In the present paper we are still not elaborating neither on the Lagrangean nor on the gauge fields for the $Z_3$-graded modification of SM; 
%this is going to be the subject of our future research.    

\newpage

\section*{Appendix I}
\vskip 0.2cm
Here we give the multiplication table of the Lie algebra spanned by $8$ generators; the six off-diagonal $Q$-matrices  $( Q_a, Q_b^{\dagger}), (a, b = 1,2,3) $ 
and the pair of diagonal matrces $(B, B^{\dagger})$. The entries correspond to the ordinary commutators $[A, B] = AB - BA$.

The overall pattern becomes clearly visible if we express all complex coefficients appearing in the table of commutators in terms of
in terms of three Greek letters. Let us introduce the following notation:
\begin{equation}
j-j^2 = \sqrt{3} \; e^{\frac{i \pi}{2}} = \alpha, \; \; \; \; j^2 - 1 = \sqrt{3} \; e^{\frac{7 i \pi}{6}} = \beta, \; \; \; \;  
1-j  = \sqrt{3} \; e^{-\frac{ i \pi}{6}} = \gamma.
\label{threealpha}
\end{equation}
 
{\small
\begin{center}
\begin{tabular}{|c|c|c|c|c|c|c|c|c|}
\hline
\raisebox{0mm}[6mm][2mm]{ \, \ \ } & $Q_1$ & $Q_2$ & $Q_3$ & $Q^{\dagger}_1$ & $ Q^{\dagger}_2$ & $Q^{\dagger}_3$ & $B$ & $B^{\dagger}$ \cr
\hline\hline
\raisebox{0mm}[6mm][2mm]{ $Q_1$ } & $0$ & $- \alpha \; Q^{\dagger}_3$ & $ \alpha \; Q^{\dagger}_2$ & $0$ & $\beta \; B^{\dagger}$ & $ - \gamma \; B$ & $ -\beta \; Q_2$ & $ \gamma \; Q_3$ \cr
\hline
\raisebox{0mm}[6mm][2mm]{ $Q_2$ } & $\alpha \; Q^{\dagger}_3$ & $ 0 $ & $-\alpha \; Q^{\dagger}_1$ & $- \gamma \; B$ & $ 0 $ & $ \beta \; B^{\dagger} $ & $- \beta \; Q_3$ & $ \gamma \; Q_1$ \cr
\hline
\raisebox{0mm}[6mm][2mm]{ $Q_3$ } & $- \alpha Q^{\dagger}_2$ & $\alpha \; Q^{\dagger}_1$ & $ 0 $ & $\beta \; B^{\dagger}$ & $- \gamma \; B $ & $ 0 $ & $- \beta Q_1$ & $ \gamma \; Q_2$ \cr
\hline
\raisebox{0mm}[6mm][2mm]{ $Q^{\dagger}_1$ } & $ 0 $ & $\gamma \; B $ & $- \beta \; B^{\dagger} $ & $ 0 $ & $- \alpha \; Q_3$ & $\alpha \; Q_2$ & $\beta \; Q^{\dagger}_3$ & $- \gamma \; Q^{\dagger}_2$ \cr
\hline
\raisebox{0mm}[6mm][2mm]{ $Q^{\dagger}_2$ } & $-\beta \; B^{\dagger} $ & $ 0 $ & $ \gamma \; B$ & $\alpha \; Q_3$ & $0  $ & $ -\alpha \; Q_1$ & $\beta \; Q^{\dagger}_1$ & $-\gamma \; Q^{\dagger}_3$ \cr
\hline
\raisebox{0mm}[6mm][2mm]{ $Q^{\dagger}_3$ } & $\gamma \; B$ & $- \beta \; B^{\dagger}$ & $ 0 $ & $- \alpha \; Q_2$ & $\alpha \; Q_1$ & $ 0 $ & $\beta \;Q^{\dagger}_2$ & $- \gamma \; Q^{\dagger}_1$ \cr
\hline
\raisebox{0mm}[6mm][2mm]{ $B$ } &$\beta \; Q_2$ &$ \beta \; Q_3$ & $\beta \; Q_1$ & $ - \beta \; Q^{\dagger}_3$ & $-\beta \; Q^{\dagger}_1$ & $ -\beta \; Q^{\dagger}_2$ & $0$& $0$ \cr
\hline
\raisebox{0mm}[6mm][2mm]{ $B^{\dagger}$ } & $- \gamma \; Q_3$ & $-\gamma \;Q_1$ & $- \gamma \; Q_2$ & $\gamma \; Q^{\dagger}_2$ & $ \gamma \; Q^{\dagger}_3$ & $ \gamma \; Q^{\dagger}_1$ & $0$ & $0$ \cr
\hline \hline
\end{tabular} 
\end{center} }
\vskip 0.2cm
{\centerline{Table 2. The commutators between eight $3 \times 3$ generators of $SU(3)$ algebra }}
{\centerline{in Kac's basis, with the coefficients $\alpha, \; \beta, \; \gamma$ given by (\ref{threealpha}).}}
\vskip 0.2cm
The multiplication table is obviously anti-symmetric, and all complex coefficients have the same absolute value $\sqrt{3}$. If we renormalize the
generators dividing every one by $\sqrt{3}$, the new generators ${\tilde{Q}}_a = Q_a/\sqrt{3}$ would satisfy the same commutator algebra with complex
renormalized structure constants ${\tilde{\alpha}} = \alpha/\sqrt{3}$, etc., with their moduli equal to $1$.

The Lie algebra defined by this table is semi-simple, what can be seen from the property that each row and each column contains {\it all six} different 
coefficients, but each of them appearing only once.

Two $3 \times 3$ blocks containing brackets between the generators $Q_a$ or $Q_a^{\dagger}$ display the same set of coefficients, 
equal to $\pm \alpha$, 
while the Cartan subalgebra generators $B$ and $B^{\dagger}$ commute (see in Table 2 the last $2 \times 2$ matrix block filled with zeros). 
The commutators of $B$ and $B^{\dagger}$ with $Q_a$'s or $Q^{\dagger}_b$'s  result in raising the index $a$
($1 \rightarrow 2, \; 2 \rightarrow 3, \; 3 \rightarrow 1$) or lowering the index $b$ ($3 \rightarrow 2, \; 2 \rightarrow 1, \; 1 \rightarrow 3$ ).

The structure of the $SU(3)$ Lie algebra is now clearly visible. It is illustrated by the root diagram (\ref{fig:Twelveroots}), displaying the third 
roots of unity ($1, j, j^2$) and third roots of $-1$ ($-1, -j, -j^2$), as well as the roots ($\pm \alpha, \pm \beta, \pm \gamma$).

\begin{figure}[hbt!]
\centering
\includegraphics[width=6.3cm, height=6.8cm]{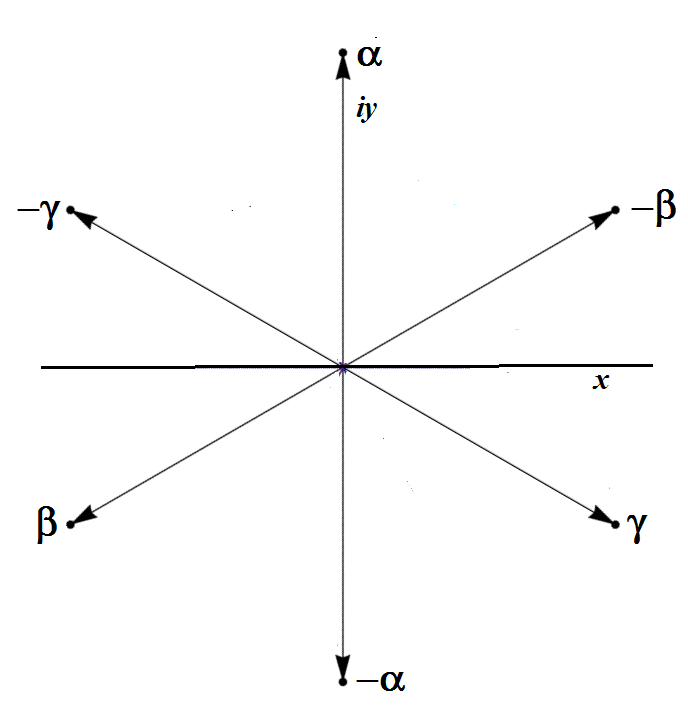}
\hskip 0.9cm
\includegraphics[width=4.7cm, height=6cm]{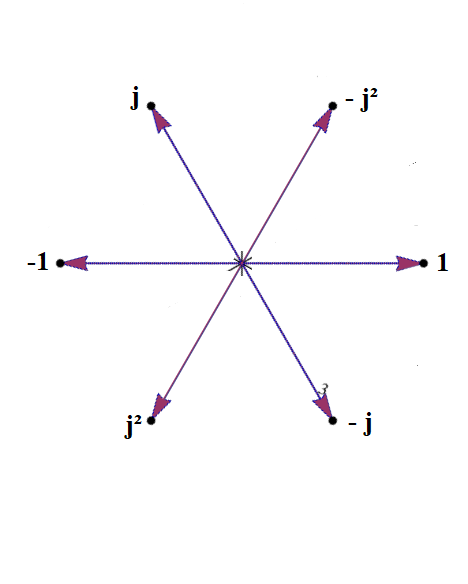}
\caption{\small{ Left: The diagram of complex coefficients $\pm \alpha, \pm \beta, \pm \gamma; |\alpha|= |\beta| = |\gamma| = 
\sqrt{3}$; Right: the sixth roots of unity} }
\label{fig:Twelveroots}
\end{figure}

\section*{Appendix II}
\vskip 0.2cm

Six traceless off-diagonal Gell-Mann matrices:
\begin{equation}
\lambda_1 = \begin{pmatrix} 0 & 1 & 0 \cr 1 & 0 & 0 \cr 0 & 0 & 0 \end{pmatrix}, \; \; \; \; 
\lambda_2 = \begin{pmatrix} 0 & -i & 0 \cr i & 0 & 0  \cr 0 & 0 & 0 \end{pmatrix}, \; \; \; 
\lambda_4 = \begin{pmatrix} 0 & 0 & 1 \cr 0 & 0 & 0  \cr 1 & 0 & 0 \end{pmatrix}, 
\label{lambda124}
\end{equation}
\begin{equation}
\lambda_5 = \begin{pmatrix} 0 & 0 & -i \cr 0 & 0 & 0 \cr i & 0 & 0 \end{pmatrix}, \; \; \; \; 
\lambda_6 = \begin{pmatrix} 0 & 0 & 0 \cr 0 & 0 & 1  \cr 0 & 1 & 0 \end{pmatrix}, 
\lambda_7 = \begin{pmatrix} 0 & 0 & 0 \cr 0 & 0 & -i  \cr 0 & i & 0 \end{pmatrix}, 
\label{lambda567}
\end{equation}
supplemented by two traceless diagonal matrices spanning the Cartan subalgebra of $SU(3)$ Lie algebra
\begin{equation}
\lambda_3 = \begin{pmatrix} 1 & 0 & 0 \cr 0 & -1 & 0 \cr 0 & 0 & 0 \end{pmatrix}, \; \; \; \; 
\lambda_8 = \frac{1}{\sqrt{3}} \;  \begin{pmatrix} 1 & 0 & 0 \cr 0 & 1 & 0 \cr 0 & 0 & -2 \end{pmatrix}. \; \; \; \;  
\label{lambda38}
\end{equation}
  
The mapping between the Cartan subalgebras, $B$ and $B^{\dagger}$ on one side and $\lambda_3$ and $\lambda_8$ on the other side,
is given by the following linear combinations:
\begin{equation}
\frac{1}{j-1} B + \frac{1}{j^2-1} B^{\dagger} = \lambda_3, \; \; \; \; - \frac{j}{\sqrt{3}} B - \frac{j^2}{\sqrt{3}} B^{\dagger} = \lambda_8, 
\label{lambda38B}
\end{equation}
or more explicitly,
\begin{equation}
% \frac{1}{j-1} B + \frac{1}{j^2-1} B^{\dagger} = 
\frac{1}{j-1} \begin{pmatrix}  1 & 0 & 0 \cr 0 & j & 0 \cr 0 & 0 & j^2 \end{pmatrix}
+ \frac{1}{j^2-1} \begin{pmatrix}  1 & 0 & 0 \cr 0 & j^2 & 0 \cr 0 & 0 & j \end{pmatrix} = 
\begin{pmatrix} 1 & 0 & 0 \cr 0 & -1 & 0 \cr 0 & 0 & 0 \end{pmatrix} = \lambda_3,
\label{BBlambda3}
\end{equation}
\begin{equation}
%- \frac{j}{\sqrt{3}} B - \frac{j^2}{\sqrt{3}} B^{\dagger} = 
- \frac{j}{\sqrt{3}} \begin{pmatrix}  1 & 0 & 0 \cr 0 & j & 0 \cr 0 & 0 & j^2 \end{pmatrix}
- \frac{j^2}{\sqrt{3}}  \begin{pmatrix}  1 & 0 & 0 \cr 0 & j^2 & 0 \cr 0 & 0 & j \end{pmatrix} = 
\frac{1}{\sqrt{3}} \begin{pmatrix} 1 & 0 & 0 \cr 0 & 1 & 0 \cr 0 & 0 & -2 \end{pmatrix} = \lambda_8.
\label{BBlambda8}
\end{equation}
The six Gell-Mann matrices (\ref{lambda124}), (\ref{lambda567}) can be expressed as linear combinations of ternary Clifford algebra generators
 $Q_a, Q^{\dagger}_{b}$ as follows:

$$\lambda_1 = \frac{1}{3} \; \left( Q_1 + Q_2 + Q_3 + Q^{\dagger}_1 + Q^{\dagger}_2 + Q^{\dagger}_3 \right),$$
$$\lambda_2 = \frac{i}{3} \; \left( Q_1 + Q_2 + Q_3 - Q^{\dagger}_1 - Q^{\dagger}_2 - Q^{\dagger}_3 \right),$$
\begin{equation}
\lambda_4 = \frac{1}{3} \; \left(j Q_1 + j^2 Q_2 + Q_3 + j^2 Q^{\dagger}_1 + j Q^{\dagger}_2 + Q^{\dagger}_3 \right),
\label{GMlambdas}
\end{equation}
$$\lambda_5 = \frac{i}{3} \; \left( j Q_1 + j^2 Q_2 + Q_3 - j^2 Q^{\dagger}_1 - j Q^{\dagger}_2 - Q^{\dagger}_3 \right),$$
$$\lambda_6 = \frac{1}{3} \; \left( j^2 Q_1 + j Q_2 + Q_3 + j Q^{\dagger}_1 +j^2 Q^{\dagger}_2 + Q^{\dagger}_3 \right),$$
$$\lambda_7 = \frac{i}{3} \; \left(j Q_1 + j^2 Q_2 + Q_3 -j^2 Q^{\dagger}_1 -j  Q^{\dagger}_2 - Q^{\dagger}_3 \right),$$

\section*{Acknowledgements}
\vskip 0.3cm
\indent

J.L. has been supported by the Polish National Science Centre (NCN) Research Project 2017/27/B/512/01902

Both authors express their thanks to Stefan Groote for valuable suggestions and remarks, and the
anonymous Referee for constructive critical comments.
One of the authors (J.L.) would like to thank Andrzej Borowiec for interesting discussion and valuable 
remarks. R.K. would like to acknowledge constructive discussions with Iwo Bia\l{}ynicki-Birula


\begin{thebibliography}{9}

\bibitem{Gaillard} The standard model of particle physics
Gaillard M K, Grannis P D and J. Sciulli F J 1999 {\it The standard model of particle physics}, {\it Rev. Mod. Phys.} {\bf 71}, S96 

\bibitem{Cottingham} Cottingham W N and Greenwood D H 2007 {\it An Introduction to the Standard Model of Particle Physics} Cambridge University Press
(second edition)

\bibitem{Bustamante} Bustamante M, Cieri L and Ellis J 2010 {\it Beyond the Standard Model for Monta\~neros}, arXiv: 0911.4409v2 

\bibitem{RKOS2014} Kerner R and  Suzuki O 2014 {\it The discrete quantum origin of the Lorentz group and the $Z_3$-graded ternary algebras},
Proceedings of RIMS Conference on Mathematical Physics, Kyoto 2013  see also https://ci.nii.ac.jp/naid/110009863886  

\bibitem{Kerner2017A} Kerner R 2017 
{\it Ternary generalization of Pauli's principle and the $Z_6$-graded algebras}, \;
 {\it Phys. Atom. Nucl.}, {\bf 80} (3) 529;  arXiv:1111.0518, arXiv:0901.3961

\bibitem{Cerejeiras} Cerejeiras P and Vaijac M B 2021 {\it Ternary Clifford Algebras} {Adv. Appl. Clifford Algebras} {\bf 31} (13) 

\bibitem{Ablamowicz} Ab{\l}amowicz R 2021 {\it On Ternary Clifford Algebras On Two Generators Defined by
Extra-Special 3-Groups of Order 27}, {\it Adv. Appl. Clifford Algebras}, to appear in 2021

\bibitem{Greensite}  Greensite J 2011 {\it An introduction to the confinement problem} Springer Lecture Notes in Physics, {\bf 821} 

\bibitem{LeeWick} Lee T D and Wick G C  1970  {\it Finite Theory of Quantum Electrodynamics}  {\it Phys. Rev. D}, {\bf 2} 1033 

\bibitem{AnselmiPiva} Anselmi D and Piva M 2017 {\it Perturbative Unitarity of Lee-Wick Quantum Field Theory}, 
 {\it Phys. Rev. D} {\bf 96} 045009 

\bibitem{Kerner2018B} Kerner R  2018 {\it Ternary $Z_2 \times Z_3$ graded algebras and ternary Dirac equation}, \; 
{\it Phys. Atom. Nucl.} {\bf 81} (6), 871; arXiv:1801.01403

\bibitem{Kerner2019B} Kerner R 2019 {\it The $Z_3$-graded extension of the Poincar\'e algebra}, arXiv:1908.02594 

\bibitem{RKJL2019} Kerner R and Lukierski J  2019 
{\it $Z_3$-graded colour Dirac equation for quarks, confinement and generalized Lorentz symmetries},
 {\it Phys. Letters B} {\bf 792} 233; arXiv:1901.10936 [hep-th] 

\bibitem{RKJL2020} Kerner R and Lukierski J 2019 {\it Towards the $Z_3$-graded approach to quarks' symmetries}, Proceedings of the $XXVI$-th conference 
on Integrable Systems and Quantum Symmetries (Prague, July $2019$), Journ. of Phys. Conf. Ser. {\bf 1416}, $012016$; arXiv 1910.05131

\bibitem{Kerner2019} Kerner R  2019  {\it The Quantum nature of Lorentz invariance}, {\it Universe}, {\bf 5} (1), 1.
https://doi.org/10.3390/universe5010001 

\bibitem{AKL2017} Abramov V, Kerner R, Liivapuu O 2017 {\it Algebras with Ternary Composition Law Combining $Z_2$ and $Z_3$ Gradings},
in: Proceedings of the International Conference on Stochastic Processes and Algebraic Structures, pp. 13-45 Springer

\bibitem{Kerner2018A} Kerner R 2018 {\it Ternary generalizations of graded algebras with some physical applications},
in {\it Mathematical Structures and Applications}, Revue Roumaine de Math\'ematiques Pures et Appliqu\'ees,
{\bf 63} (2), pp. 107-141 

\bibitem{Cayley} Cayley, A  1858  {\it Phil. Trans. Royal Soc. of London} {\bf 148}, pp. 17-37 

\bibitem{Sylvester} Sylvester, J. J  1884 {\it Johns Hopkins University Circulars} I: 241-242; ibid II (1883) 46; ibid III pp 7-9 

\bibitem{RKOS} Kerner R  Suzuki O 2012 {\it  Int. J. Geom. Methods Mod. Phys.} {\bf 09}, 1261007

\bibitem{SylvesterA} J. J. Sylvester J J (1882) {\it A word on nonions}, John Hopkins Univ. Circ. {\bf 1}, 241-242; 
ibid, (1883) {\bf 46} (E).  (in {\it The Collected Mathematical Papers of James Joseph Sylvester}, (Cambridge Univ. Press, Cambridge, 1909), 
Vol. {\bf 3}, pp. 647-650.   

\bibitem{Kac1994} Kac  V G  1984  {\it "Infinite-Dimensional Lie Algebras" } Cambridge Univ.Press

\bibitem{Colemandula} Coleman S and Mandula J  1967 {\it All possible symmetries of the S-Matrix} {\it Phys. Rev.} {\bf 159}, 1251

\bibitem{Raifeartaigh} O'Raifeartaigh L 1965  {\it Lorentz Invariance and Internal Symmetry}, {\it Phys.Rev.} {\bf 139}, B1052 

\bibitem{Sogami2019} Sogami I 2018 {\it Unified description of quarks and leptons in a multi-spinor field formalism }, arXiv:1512.09283, see also 

\bibitem{Sogami2020} Sogami I and  Kozumi K 2020 {\it Renovation of the Standard Model with Clifford-Dirac algebras for chiral-triplets},
{\it Phys. Letters B}, {\bf 808}, 135622.

\bibitem{PatiSalam} Pati J C and Salam A 1974 {\it Phys Rev. D} {\bf 10}  p.275 

\bibitem{MarschNarita} Marsch E and Narita Y 2020 {\it The European Physical Journal Plus}, {\bf 135}, (10) Article number: 782 

\bibitem{Finkelstein} Finkelstein D, Finkelstein S R, Holm C 1986 {\it Hyperspin manifolds}, {\it Int. Jour,. of Theor. Phys.}, 
{\bf 25}, pp. 441-463. 

\bibitem{Sitarz} Paschke M, Scherk F and Sitarz A 1999 {\it Can (non-commutative) geometry accomodate lepto-quarks?} {\it Phys. Rev. D54}
035003

\bibitem{Todorov} Todorov I and Dubois-Violette M 2018 {\it Deducing the symmetry of Standard Model from automorphism and structure
groups of the exceptionla Jordan algebra}, {\it Int. Journ. Mod. Phys. A33} 1850118

\bibitem{Bochniak} Bochniak A and Sitarz A 2020 {\it A spectral geometry for the Standard Model without the fermion doubling}
arXiv: 2001.02902

\end{thebibliography}
\end{document}